\documentclass[preprint,11pt,table]{elsarticle}

\usepackage{algorithmic}
\usepackage{graphicx}
\usepackage{textcomp}
\usepackage[table]{xcolor}
\usepackage{todonotes}

\usepackage{makecell}
\usepackage{booktabs}
\usepackage{pifont}% http://ctan.org/pkg/pifont
\usepackage{array,multirow,graphicx}
\newcommand{\down}{\ding{216}}%
\newcommand{\up}{\ding{218}}%

\usepackage{color,soul}
\definecolor{Red}{rgb}{0.5,0.0,0.0}
\definecolor{Green}{rgb}{0.0,0.5,0.0}
\definecolor{Blue}{rgb}{0.0,0.0,0.8}

\newcommand{\red}[1]{#1}
\newcommand{\green}[1]{#1}
\newcommand{\blue}[1]{#1}

\usepackage[frozencache,cachedir=.]{minted}
\usepackage{caption}

\usepackage{subcaption}
\usepackage[many,minted]{tcolorbox}% version 3.03 or better
\usepackage{hyperref}

\newcommand{\ra}[1]{\renewcommand{\arraystretch}{#1}}

\usepackage{amssymb}
\journal{Journal of Systems and Software}

\begin{document}

\begin{frontmatter}

%% Title, authors and addresses

%% use the tnoteref command within \title for footnotes;
%% use the tnotetext command for theassociated footnote;
%% use the fnref command within \author or \address for footnotes;
%% use the fntext command for theassociated footnote;
%% use the corref command within \author for corresponding author footnotes;
%% use the cortext command for theassociated footnote;
%% use the ead command for the email address,
%% and the form \ead[url] for the home page:
%% \title{Title\tnoteref{label1}}
%% \tnotetext[label1]{}
%% \author{Name\corref{cor1}\fnref{label2}}
%% \ead{email address}
%% \ead[url]{home page}
%% \fntext[label2]{}
%% \cortext[cor1]{}
%% \affiliation{organization={},
%%   addressline={},
%%             city={},
%%             postcode={},
%%             state={},
%%             country={}}
%% \fntext[label3]{}

\title{Piloting Copilot, Codex, and StarCoder2: Hot Temperature, Cold Prompts, or Black Magic?}

%% use optional labels to link authors explicitly to addresses:
%% \author[label1,label2]{}
%% \affiliation[label1]{organization={},
%%             addressline={},
%%             city={},
%%             postcode={},
%%             state={},
%%             country={}}
%%
%% \affiliation[label2]{organization={},
%%             addressline={},
%%             city={},
%%             postcode={},
%%             state={},
%%             country={}}

\author[ens,irisa]{Jean-Baptiste Döderlein}
\ead{jean-baptiste.doderlein@ens-rennes.fr}

\author[univrennes,cnrs,irisa]{Nguessan Hermann Kouadio}
\ead{nguessan-hermann.kouadio@inria.fr}

\author[univrennes,IUF,cnrs,irisa]{Mathieu Acher}
\ead{mathieu.acher@irisa.fr}

\author[cnrs,univrennes,irisa]{Djamel Eddine Khelladi}
\ead{djamel-eddine.khelladi@irisa.fr}

\author[univrennes,cnrs,irisa]{Benoit Combemale}
\ead{benoit.combemale@irisa.fr}

\affiliation[ens]{organization={ENS Rennes},
  city={Bruz},
  country={France}
}

\affiliation[univrennes]{organization={Univ Rennes},
  city={Rennes},
  country={FR}
}

\affiliation[IUF]{%
  organization={IUF},
  city={Rennes},
  country={FR}
}

\affiliation[cnrs]{%
  organization={CNRS},
  city={Rennes},
  country={FR}
}

%\affiliation[inria]{%
%  organization={Inria},
%  city={Rennes},
%  country={FR}
%}

\affiliation[irisa]{%
  organization={IRISA},
  city={Rennes},
  country={FR}
}

\begin{abstract}

Language models are promising solutions for tackling increasing complex problems. In software engineering, they recently gained attention in code assistants, which generate programs from a natural language task description (prompt). They have the potential to save time and effort but remain poorly understood, limiting their optimal use.
In this article, we investigate the impact of input variations on two configurations of a language model, focusing on parameters such as task description, surrounding context, model creativity, and the number of generated solutions. We design specific operators to modify these inputs and apply them to three LLM-based code assistants (Copilot, Codex, StarCoder2) and two benchmarks representing algorithmic problems (HumanEval, LeetCode). Our study examines whether these variations significantly affect program quality and how these effects generalize across models.

Our results show that varying input parameters can greatly improve performance, achieving up to 79.27\% success in one-shot generation compared to 22.44\% for Codex and 31.1\% for Copilot in default settings. Actioning this potential in practice is challenging due to the complex interplay in our study—the optimal settings for temperature, prompt, and number of generated solutions vary by problem.

Reproducing our study with StarCoder2 confirms these findings, indicating they are not model-specific. We also uncover surprising behaviors (e.g., fully removing the prompt can be effective), revealing model brittleness and areas for improvement.

Overall, this work opens opportunities to envision (automated) strategies for enhancing performance of language model-based code assistants, but also questions their reliability and robustness.

\end{abstract}

%%Research highlights
%\begin{highlights}
%\item Research highlight 1
%\item Research highlight 2
%\end{highlights}

\begin{keyword}
%% keywords here, in the form: keyword \sep keyword
language models \sep code generation \sep programming tasks
%% PACS codes here, in the form: \PACS code \sep code
%\PACS 0000 \sep 1111
%% MSC codes here, in the form: \MSC code \sep code
%% or \MSC[2008] code \sep code (2000 is the default)
%\MSC 0000 \sep 1111
\end{keyword}

\end{frontmatter}

%% \linenumbers

%% main text
\section{Introduction}

Language models are gaining momentum and capable of tackling more and more problems from linguistics, maths, commonsense reasoning, biology, physics, etc. 
BERT \cite{BERT}, GPT-2 \cite{GPT2}, GPT-3 \cite{GPT3}, PaLM \cite{chowdhery2022palm}, to name a few, are scaling to support a variety of tasks such as text generation, question-answering, text classification, arithmetic on numbers, and many others \cite{srivastava2022beyond,du2021glam,rush-etal-2015-neural,jiang2020can,geva-etal-2020-injecting}. 
In software engineering, code assistants based on language models have been proposed and are now deployed at scale for supporting programmers, such as GitHub Copilot \cite{github-copilot}. 
Based on \emph{prompts}, composed of both the \emph{description of a programming task} written in natural language and the surrounding \emph{context} (e.g., existing code, function signatures, targeted language, cursor, authors, shebang), programs are automatically written in a given programming language (Python, Java, C++, etc.). 
The promise is to provide a comprehensive working solution or a set of candidate programs for a given programming task. 
Tools like Copilot or Codex hence have the potential to save time and effort when writing code. 

However, the strengths and weaknesses of these systems are currently poorly understood, preventing them from being used optimally. On the one hand, there are impressive demonstrations, showing the ability to produce programs on non-trivial programming problems or tasks. 
But there also is the nagging assumption that these systems are simply reciting code that is already on the Internet (e.g., Github). 
Furthermore, early studies suggest that the quality of solutions appears to vary greatly in some problems and targeted programming languages~\cite{nguyen-empirical-2022}. These assistants seem in particular sensitive to the way the developer communicates and interacts. 

Given a programming task, developers can pilot/drive code assistants in different directions to achieve their goal. They can vary the prompt, for example, formulate the programming task differently or change the context. They can also change other language models input parameters, such as augmenting the creativity of the assistant (through the \emph{temperature} of language models), or change the number of expected solutions that are eventually proposed. 
With high flexibility, developers can communicate at a high level of abstraction, in a declarative way, focusing on the goal rather than the how. 
The counterpart is that the specification might be brittle and not properly or systematically understood by code assistants. 
There is also the question of what strategy to choose when developers try to find a solution: changing some terms in the programming task description. Changing the signature of the function? Augmenting or decreasing the temperature? Etc.
There are anecdotes here and there about "language model engineering" (e.g., prompt engineering), but this has not been systematically studied in the context of code assistants. 

In this article, we hypothesize that variations of input parameters of language models (e.g., prompts and temperatures) on the same problem can have a significant impact on the quality of the generated programs. These variations can be leveraged to (1) assess and understand the sensitivity (or robustness) of code assistants, hence, their potential and limitations; (2) envision (automated) strategies for improving performance. % developers to effectively communicate. 

To do so, we first design and develop a set of operators to automatically vary the input parameters. These operators can remove, augment, or simply rewrite an original programming task description, as well as varying the context and other input parameters such as temperature and the number of expected solutions.  
The idea is to feed code assistants with different variations, observe the effects on generated programs, and possibly better understand the impact of the input parameters on the resulting performance of the language model. Performance is defined in this article as the ability or not to find at least one solution among the proposed ones that pass all the test cases.

%address the following research questions:
%% Given a variation, to what extent are code assistants sensitive? 
% How sensitive are code assistants to variations of prompts and temperature?
%What are the most effective variations for degrading or improving the %accuracy of code assistants? 
%% Can we recommend to developers a systematic strategy and set of guidelines %when piloting code assistants? 
%How much can we improve performance of code assistants when varying prompt %and temperature? 
We conducted a study that considers two code assistants, namely Copilot and Codex. Our study leverages two datasets, namely HumanEval and LeetCode mostly representing algorithmic problems, as well as our set of operators.  
% \newtext{Numerous works~\cite{drori-solving-2021,tang-solving-2021,sobania-choose-2021,nguyen-empirical-2022, ziegler-productivity-2022, vaithilingam-expectation-2022,mastropaolo2023robustness} share this concern. 
% This line of empirical research is not meant to reach a final conclusion. Its aim is to understand the current potential and limitations of the subject, apart from any hype or buzz.}
% Obviously, the goal of this line of empirical research is not to draw a definitive conclusion, but rather to gain insights about their current potential, beyond the hypes and buzz}.
Our experiments span %numerous programming tasks (446 problems) 
446 programming tasks as problems
with different difficulties and six programming languages. 
We also vary the number $k$ of code samples generated per task, from $k=1$ (one shot) to $k=100$. 
We study the sensitivity of code assistants and the effectiveness of our variations in different settings and usage scenarios. 
Our experiments have been executed around the period of August 2022. 

Owing to the current trend, we are well aware that new technologies based on language models will continuously emerge, e.g., Copilot X and ChatGPT4~\cite{openai2023gpt4} have been recently released. By construction, this line of empirical research is not meant to reach a final conclusion. Its aim is to understand the current potential and limitations of the subject, apart from any hype or buzz. Further studies can confirm or contradict our insights, e.g., about the prompt or temperature sensitivity. Interestingly, the training set of Codex does not contain programs (i.e., solutions) of HumanEval since, by construction, HumanEval was precisely designed to evaluate Codex with an external dataset not used during the training~\cite{chen-evaluating-2021}. It is not the case of recent language models like GPT4 that are subject to contamination~\cite{openai2023gpt4}. \textbf{Hence, the period of our experiments retrospectively gives us a unique opportunity to understand two code assistants in a setting where the contamination was less severe and where simply reciting solutions could well explain high performance of some language models.}
To the best of our knowledge, very few works study the sensitivity and robustness of language model-based code assistants. An exception is the work of Mastropaolo et al.~\cite{mastropaolo2023robustness} published at ICSE'23 that asked Copilot to automatically generate Java methods starting from variations of Javadoc description. We pursue a similar objective, but there are several differences. We consider two related yet differently configured language model-based assistants (Copilot and Codex). Copilot can be thought of as a thorough engineering of Codex, providing a chance for research to evaluate sensitivity in various configuration settings and acquire insights on how to pilot language models. From this regard, we vary the temperature and $k$, apply different variations of prompts and target programs written in different programming languages. 
\red{Finally, we reproduce our original study on the open-source LLM StarCoder2.}

% adding new tests as variants of existing ones

Our contributions can be summarized as follows.
\begin{itemize}
    \item The design and development of a set of operators for automatically varying language models input parameters. The inspiration of our work both comes from software testing techniques (e.g., mutation testing, test amplification as variants of existing ones) and recent advances in language models for tuning prompts%~\cite{shin_autoprompt_2020,gehman-etal-2020-realtoxicityprompts} ; 
    \item The design of a study over \red{three} code assistants and two benchmarks. Prior studies considered a limited number of problems, programming languages, and configurations of code assistants. We are also unaware of works that leverage prompt variations and temperatures' values in the context of code generation. 
    \item The analysis of results that demonstrate that varying input parameters can significantly improve the performance of language models. However, there is a tight dependency when varying the temperature, the prompt and the number of generated solutions, making potentially hard for developers to properly control the parameters to obtain an optimal result.
\end{itemize}

\section{Background and Motivation}

% This section introduces the necessary background and a motivating example. 

\subsection{Language models and Code Suggestions}
% \subsection{Codex, Copilot}

A language model (LM) is a probabilistic model over natural language. In practice, from a given prompt, a language model provides a set of results. In software engineering, language models such as Codex are called from a prompt including a description of a programming task in natural language, as well as all the surrounding context. Such a context is composed of information such as the existing code (e.g., imports...), the offset of the cursor, the targeted language, the authors, the shebang...

Beyond the prompt, language models might be tuned regarding specific parameters. In particular, the \emph{temperature} hyperparameter controls the creativity of the language model. 
The temperature value typically varies from 0 to 1. The temperature is used to adjust the distribution of the model's predicted next word: higher values lead to more diverse and unpredictable outputs, and lower values lead to more conservative, predictable outputs.
Another parameter is the expected number $k$ of generated solutions. The principle is to sample from the $k$ most probable programs. 
When $k$ equals to 1, only one program is generated and is hopefully a valid solution. 

\subsection{Prompts}

When using Copilot and Codex to generate code, the user provides a context to the model. This could be a text in natural language or some pieces of existing code.

The prompt is the text/code that the model needs to complete. This includes the comment (e.g., the docstring of the function in Python) but also the function signature. The function signature consists of the function name, the number of arguments, their name, and also in type in the case of typed languages.

\emph{Prompt sensitivity: An example.}
% This section introduces a motivating example of code suggestions with input prompts. 
 When using language model-based assistant tools to generate code, there are several ways to express what should be completed. The most common is to use the comment of the function.
But there are multiple ways to phrase the prompt. One can change the prompt while keeping the same meaning. Yet, sometimes, a small variation could drastically change the model performance.

In Figure~\ref{fig:codeExample1} we use an example of a prompt from HumanEval \cite{chen-evaluating-2021} dataset. The function takes two integers and returns the biggest even number in the range of the two integers, $-1$ if no even number is found. With Copilot, the model fails to provide a correct answer (it returns the smallest even number instead). However, the model provides a correct answer when we add the sentence "Write a quick algorithm to solve this problem." at the end of the prompt. This sentence does not provide any additional information to the model about the problem or the algorithm to solve the problem, but it improves the model's output. Furthermore, we observe that if we further modify the prompt by removing the examples, the result is not correct.

These examples suggest that code generation is sensitive to prompt variation, and that modification of the prompt can cause the model to provide a correct or wrong answer within different contexts.

\begin{figure*}
  \begin{subfigure}[t]{0.33\textwidth}
    \centering
    \tcbset{
    width=(\linewidth+2pt),
    boxsep=0pt,
    left=3pt,
    right=0pt,
    top=3pt,
    bottom=1pt,
    colback=blue!4,
    colframe=blue!4,
    boxrule=0pt,frame hidden,
    arc=0pt,
    outer arc=0pt,
    }
    \begin{tcolorbox}
    \begin{minted}[fontsize=\scriptsize]{python}
def choose_num(x, y):
    """
    This function takes two 
    positive numbers x and y 
    and returns the biggest 
    even integer number that 
    is in the range [x, y] 
    inclusive. If there's no 
    such number, then the 
    function should return -1.
    
    For example:
    choose_num(12, 15) = 14
    choose_num(13, 12) = -1
    """
    \end{minted}
    \end{tcolorbox}
    \tcbset{
    size=minimal,
    width=(\linewidth+2pt),
    boxsep=0pt,
    left=3pt,
    right=0pt,
    top=0pt,
    bottom=2pt,
    colback=green!4,
    colframe=green!4,
    boxrule=0pt,frame hidden,
    arc=0pt,
    outer arc=0pt,
    enlarge top initially by=-2mm
    }
    \begin{tcolorbox}
    \begin{minted}[fontsize=\scriptsize]{python}
    if x % 2 != 0:
        x += 1
    if y % 2 != 0:
        y -= 1
    if x > y:
        return -1
    if x % 2 == 0:
        x += 1
    while x <= y:
        if x % 2 == 0:
            return x
        x += 2
    return -1
    \end{minted}
    \end{tcolorbox}
    \caption{Original prompt} \label{fig:codeExample1:original}
\end{subfigure}\hfill
  \begin{subfigure}[t]{0.33\textwidth}
  \centering
  \tcbset{
    width=(\linewidth+2pt),
    boxsep=0pt,
    left=3pt,
    right=0pt,
    top=3pt,
    bottom=1pt,
    colback=blue!4,
    colframe=blue!4,
    boxrule=0pt,frame hidden,
    arc=0pt,
    outer arc=0pt,
    }
    \begin{tcolorbox}
    \begin{minted}[fontsize=\scriptsize]{python}
def choose_num(x, y):
    """
    This function takes two 
    positive numbers x and y 
    and returns the biggest 
    even integer number that 
    is in the range [x, y] 
    inclusive. If there's no 
    such number, then the 
    function should return -1.

    For example:
    choose_num(12, 15) = 14
    choose_num(13, 12) = -1
    
    Write a quick algorithm to 
    solve this problem.
    """
    \end{minted}
    \end{tcolorbox}
    \tcbset{
    size=minimal,
    width=(\linewidth+2pt),
    boxsep=0pt,
    left=3pt,
    right=0pt,
    top=0pt,
    bottom=2pt,
    colback=green!4,
    colframe=green!4,
    boxrule=0pt,frame hidden,
    arc=0pt,
    outer arc=0pt,
    enlarge top initially by=-2mm
    }
    \begin{tcolorbox}
    \begin{minted}[fontsize=\scriptsize]{python}
    if x % 2 == 0:
        x = x
    else:
        x = x + 1
    if y % 2 == 0:
        y = y
    else:
        y = y - 1
    if x > y:
        return -1
    else:
        return max(x, y)
    \end{minted}
    \end{tcolorbox}
    \caption{Modified prompt 1} \label{fig:codeExample1:mod1}
    \end{subfigure}\hfill
\begin{subfigure}[t]{0.33\textwidth}  
  \centering
  \tcbset{
    width=(\linewidth+2pt),
    boxsep=0pt,
    left=2pt,
    right=0pt,
    top=3pt,
    bottom=1pt,
    colback=blue!4,
    colframe=blue!4,
    boxrule=0pt,frame hidden,
    arc=0pt,
    outer arc=0pt,
    }
    \begin{tcolorbox}
    \begin{minted}[fontsize=\scriptsize]{python}
def choose_num(x, y):
    """
    This function takes two 
    positive numbers x and y 
    and returns the biggest 
    even integer number that 
    is in the range [x, y] 
    inclusive. If there's no 
    such number, then the 
    function should return -1.
    """
    \end{minted}
    \end{tcolorbox}
    \tcbset{
    size=minimal,
    width=(\linewidth+2pt),
    boxsep=0pt,
    left=2pt,
    right=0pt,
    top=0pt,
    bottom=2pt,
    colback=green!4,
    colframe=green!4,
    boxrule=0pt,frame hidden,
    arc=0pt,
    outer arc=0pt,
    enlarge top initially by=-2mm
    }
    \begin{tcolorbox}
    \begin{minted}[fontsize=\scriptsize]{python}
    if x % 2 == 1:
        x += 1
    if y % 2 == 0:
        y -= 1
    for i in range(x,y+1,2):
        if x <= i <= y:
            return i
    return -1
    \end{minted}
    \end{tcolorbox}
    \caption{Modified prompt 2} \label{fig:codeExample1:mod2}
    \end{subfigure}
  \caption{Code generation of HumanEval problem \rm \tt choose\_num \tt \rm with original prompt (a) and modified prompts (b and c). There is the prompt (blue background) and the code generated by Copilot (green background)}
  \label{fig:codeExample1}
\end{figure*}

\section{Experimental set up}

This section describes our experimental setup to explore the effect of prompt variations and temperature parameter on the performance of language model-based code assistants.

\subsection{\red{Three} code assistants, two configurations of one language model \red{and an open-source model}}%\todo{\red{should we add starcoder to the story here ?}}
Several language models and code assistant exist for code suggestion. 
We first consider two popular code assistants, Codex and Copilot. 
There are several reasons why we chose these two tools: they provide state-of-the-art performance \red{at the time of this experiment in 2022}; they are capable of targeting different programming languages; they are rather mature tools, already widely used, and come with a set of APIs. 
Another interesting property is that Codex and Copilot rely on the same language model. 
This apparent similarity in fact hides many differences that we wish to explore.
\emph{Copilot can be thought of as a thorough engineering of Codex, providing a chance for research to evaluate sensitivity in various configuration settings and acquire insights about how to configure (pilot) language models.}
% research on the sensitivity of AI code generation tools is a new topic that has never been done before.
Note that our experiment was run in the summer of 2022, which influenced our choice of tools. In particular, we used the available version of Copilot at that time and not today version of Copilot X. We also were able to experiment with Codex that was available with open access at that time. Knowing about the contamination issues with GPT-4 on which Copilot X is based, our paper is at unique position of studying two tools that were not contaminated with HumanEval programs. \red{Finally, we replicated our experiment with the open-source LLM StarCoder in 2024.} 

\subsubsection{Copilot}

Copilot has been built on top of one of the Codex language model as a code assistant with fixed parameters (e.g., temperature) and an associated development tool that is available from Github as an IDE extension. Copilot is described as an ``AI pair programmer`` at offers code suggestions in real time \cite{github-copilot}. It can produce code in different ways: It can write code from comments, write tests for code that is already written, or complete code that is being written like tools like Intellisens \cite{intellisense} does. 

The focus here is on generating code from comments (and the function signature). In regular use, Copilot communicates in real time about all the changes made on the document and sends additional information about the context (document name, path, cursor position, language, etc.). Previous studies evaluating the performance \cite{nguyen-empirical-2022} or security \cite{pearce-asleep-2021} of the code produced by Copilot retrieved the results for each prompt, manually from VS Code. In order to increase the speed of the tests and to significantly increase the number of samples evaluated, we directly made calls to the backend of the Visual Studio Code module. The user is authenticated once manually from the Neovim extension, then the calls are made to Copilot executed using NodeJS with LSP. When used in the IDE, the Copilot extension does not simply make requests, but builds contexts during generation and decides according to the context if a request is needed \cite{copilot-reverse}. However, as we use a primitive in the LSP to request a query, no context other than the name of the file and programming language are provided.

In the Copilot evaluation, we restricted ourselves to the default parameter (temperature, top\_p). Although these parameters seem to be modifiable \cite{pearce-asleep-2021}, the default values are not known, and the default parameters better reflect the codes proposed to the end user. Moreover, we did not use the possibility of the panel mode allowing generating several codes. Indeed, the codes retrieved by the backend sometimes correspond to the code of the whole file and sometimes to the simple completion of the code, thus preventing a good automation of the process. Overall, all Copilot codes are evaluated in one-shot.

\red{Since 2022, Copilot has evolved significantly, including changes to the underlying LLM it uses. In the remainder of this study, "Copilot" refers specifically to the version available during our experiments conducted in the summer of 2022. This historical perspective is valuable, as it provides insights into a version of Copilot that was free from HumanEval contamination and operated under a specific, now-documented configuration. These conditions offer a unique opportunity to analyze its behavior before more recent developments introduced potential dataset contamination and modifications in model tuning. \red{Furthermore, comparing these past results with more recent ones—such as those obtained with StarCoder2—allows us to assess whether key observations remain consistent over time and across different models, helping to better understand the evolving landscape of LLM-based code generation.}}

\subsubsection{Codex}

Codex, more precisely code-davinci-002, is a language model developed by OpenAI and available in private beta at the time of writing. It is part of the same family as the model used by Copilot. Contrary to Copilot, it has an API that facilitates the automation of the evaluation of produced codes. The API offers many parameters like: temperature, top\_p, number of code produced. For the evaluation, OpenAI advises not to change the temperature and top\_p at the same time, so we set top\_p to 1 (default parameter) and then performed the tests for the temperatures $0.0, 0.2, 0.4, 0.6, 0.8, 1$ \cite{OpenAI2024}. Interestingly, given a prompt, Codex can generate more than one program and this feature is well-supported.

% To compare directly with Copilot, we used $pass@1$, but we also made measurements for $pass@10$ and $pass@100$ in our experiments.

\red{
\subsubsection{StarCoder2}

Since we originally evaluated Codex and Copilot in 2022, we replicated our experiment with StarCoder2, a more recent open-source model, to assess whether our findings hold across different LLMs. Specifically, we used StarCoder2-15B, a 15-billion-parameter model released in February 2024, trained on 600+ programming languages from The Stack v2 dataset~\cite{starcoder2}.

Like Codex, StarCoder2 is free from contamination with some benchmarks (e.g., HumanEval), ensuring a fair evaluation of its code generation capabilities. Its fully open-source nature provides additional transparency, offering valuable insights into its training data and architecture. Beyond its openness, StarCoder2 demonstrates strong code generation performance, making it a relevant model to study. Its effectiveness, however, remains influenced by key input parameters such as temperature and prompt variations. By including StarCoder2 in our study, we aim to determine whether similar parameter variations impact its performance and how these factors can be managed. % leveraged to optimize code generation.
}

\subsection{Datasets}
\label{subsec:datasets}
We selected two datasets, namely HumanEval and Leetcode. 

\subsubsection{HumanEval}
HumanEval is a dataset originally proposed in~\cite{chen-evaluating-2021}, composed of 146 codes to complete written in Python. Each function is given a function name, its parameters and a docstring indicating the purpose of the function (see Fig \ref{fig:codeExample1}). Moreover, each function has a test suite. % \red{This dataset is only made up of handwritten problems avoiding recitation, e.g., cloning code from the learning set.}
This dataset is only composed of handwritten problems thus mitigating recitation, e.g., cloning the code of the training set. As this test set was designed and created to evaluate Codex itself, no code in this dataset is present in the training set of Codex and therefore Copilot.

\subsubsection{Leetcode}
\label{subsec:leetcode}

LeetCode is an online platform that provides a collection of problems to help software engineers prepare for technical interviews. 
It offers a variety of problems in different areas and programming languages. 
Nguyen et al. originally considered 33 problems in 4 languages~\cite{nguyen-empirical-2022}. We considered a larger dataset: we compiled 300 problems from Leetcode, chosen among the most liked problems and equitably distributed by difficulty (100 Easy, 100 Medium, 100 Hard). Each problem is present in 6 languages (Python 3, Javascript, Java, C, C++, C\#).
Problems that could not be tested in any of the 6 languages, and problems requiring multiple functions were discarded. Each problem is given as a comment at the beginning of the file corresponding to a short description of the problem, and the signature of the function to be implemented. An additional comment is sometimes provided to indicate the implementation of the elements being passed in the signature (e.g., binary trees). 

The tests used by Leetcode are not public, so the problems have been submitted to Leetcode through the GraphQL API.
The API naturally exhibit rate limits, and it is hardly possible to check, e.g., 100 candidate programs for all problems. 
We thus decided to consider Leetcode in one-shot (only one program can be submitted), but in return we did explore the variations of temperature and prompt.

Unlike HumanEval, this dataset is subject to a risk of recitation because many Github repositories contain solutions to some of the problems and may be part of the training set~\cite{copilot-recitation}. 
There are not necessarily solutions to all the Leetcode problems and in all the programming languages.
Even if a solution is available, sensitivity to prompts or temperature remains an interesting motivation. Finally, although the issue of data contamination with LeetCode may arise, it has not been proven that the model is unaffected by the prompt, particularly in generating tokens previously encountered during training.

\subsubsection{Rationale}

Copilot and Codex have similarities, but also important differences. 
On the one hand, Copilot is treated as a black box applying modifications and an unknown choice of parameters and treatments. We do not know which model and exact parameters are used, though some technical insights have been reported in the grey literature~\cite{copilot-reverse}. 
On the other hand, Codex, is a raw language model that offers more freedom in the choice of parameters. 
It is thus an opportunity to explore the configuration space and the potential of a language model. 
This choice allows us to compare the results of a tool already \emph{engineered} and configured (Copilot) against the results of the raw model (Codex) it uses. 
\red{StarCoder2 extends this comparison by introducing a fully open-source alternative, where both model architecture and training data are transparent, allowing for a more controlled and reproducible study of input parameter variations and their impact on code generation performance.}

\subsection{Variables and metrics}
To generate one or more code suggestions from a language model $LM$, we feed it with a prompt $P$, a temperature $T$ and a number of code to be generated $K$, such as $LM(P, T, K) \longrightarrow \{code*\}$.

The prompt $P$ is composed of the function signature, the description of the programming task (also known as documentation), and the context of the prompt. The context is composed of all information given at the time the language model is used. This includes the existing code (e.g., imports), the offset of the cursor, the targeted language, the authors, the shebang, among other factors.
%The context contains the information on the requested language (e.g., python, java, etc.). 

Our experiment aims to measure the influence on the suggested code when varying the prompt and the temperature to the language models in contrast to when it was not. This was the independent variable we controlled, i.e.
prompt variation and temperature variation. To measure their effects, we observed one dependent variable, namely the performance of the language models without any variation and with variations.
We define the performance of a language model as follows:

$$LM_{performance} = \frac{problemSolved}{totalProblems}$$

$problemSolved$ is the number of problems that a language model was able to generate at least one correct answer to from $\{code*\}$, i.e., a code suggestion compiling and passing the suite test of a given problem.
We relied on the test suites that come with the different datasets.

 We considered different scenarios:
\begin{itemize}
    \item $k$ equals to 1 (one-shot) where only one candidate program is considered. Copilot is necessarily in this case for both HumanEval and Leetcode. Although Codex can generate more than one program, we consider a one-shot scenario for Leetcode due to API restrictions for checking programs and as in the work by Nguyen et al.~\cite{nguyen-empirical-2022}. The one-shot scenario is typical of developers' workflow that want a given solution and do not want to review alternative solutions. 
    \item $k$ equals to 10 (resp. 100) where 10 (resp. 100) candidate programs are considered. Only Codex can be used as such for the HumanEval dataset.  
    \end{itemize}
We used the pass@k metric, that, given $k$ submitted solutions to a problem, considers a problem to be successful if one of the solutions passes all the tests~\cite{kulal-spoc-2019}. Specifically, we computed pass@1, pass@10 and pass@100 using the unbiased estimator defined by Chen et al.~\cite{chen-evaluating-2021} where $k$ solutions are randomly picked from $n$ samples:

\begin{equation}
\text{pass@k := }\: \underset{\text{Problems}}{\mathbb{E}}
\left [ 1 - \cfrac{\binom{n-c}{k}}{\binom{n}{k}} \right ]
\end{equation}

For Leetcode, we only used pass@1 for Codex and Copilot ($k=1$ and $n=1$). 
For the codes generated with Codex on HumanEval, we generated 100 solutions for each problem (i.e., we fix $n=100$) and then report on pass@1, pass@10, and pass@100.

\subsection{Research Questions}\label{RQs}

We define the following research questions.

\newcommand{\rqOne}[0]{What is the performance of Copilot and Codex over HumanEval and Leetcode?} 
\newcommand{\rqTwo}[0]{What is the impact of prompt variation on Copilot and Codex performance?}
\newcommand{\rqThree}[0]{What is the impact of the variation of both temperature and prompts on Codex performance?}
\newcommand{\rqFour}[0]{What are the best prompt variations and temperatures for Codex to maximize its performance over a given set of problems?}
\newcommand{\rqFive}[0]{How much can we improve the performance of Codex by tuning the temperature and prompts \emph{per} a specific problem?}

\begin{itemize}
    \item[RQ1] \emph{\rqOne}~
    We use the original prompts and the default temperature of Copilot and Codex. 
Hence, unlike the following research questions, we do not make vary the prompts or the temperature. 
Our goal is to establish a first trend and a baseline for the performance of the two tools that use the same language model.
We also aim to gather insights about performance sensitivity over datasets (Leetcode vs HumanEval) and programming languages (e.g., Python vs C).
    \item[RQ2] \emph{\rqTwo~}
    Our goal is to assess the sensitivity of Copilot and Codex with respect to prompt modification.  
    We make vary the prompts, leveraging different operators acting over the programming task description, and some information from the surrounding context (e.g., function signature). However, we rely on default temperature of Copilot and Codex.
    \item[RQ3] \emph{\rqThree~}
    We aim to establish the impact of varying all input parameters, including temperature, prompt, and number $k$ of generated solutions. % We conduct a multi-dimensional study to analyze the optimal combination. 
    \item[RQ4] \emph{\rqFour~}
    % \todo{} 
    We aim to identify the best parameters for Codex to maximize its performance for a given set of problems (on average). 
    \item[RQ5] \emph{\rqFive~}
    We aim to identify the best parameters for Codex to maximize its performance for a given problem. Intuitively, the best parameters on average are not necessarily the best ones for a specific problem. 
    % We \todo{do something}
\end{itemize}

\section{Prompt Variations Approach}

\begin{table}\centering
\ra{1.2}
\begin{tabular}{@{}cl@{}}
\toprule
Categories & Modification \\
\midrule
\multirow{12}{*}{\makecell[tc]{Documentation \\ modification}} & Original  \\ 
& No Documentation  \\
& No Examples  \\
& Chain of Thought prompt \\
& \textit{- "Algorithm :"}  \\
& \textit{- "Complexity"}  \\
& \textit{- "Let's thinks step by step :"}  \\
& Additionnal instruction \\
& \textit{- "Write a quick algorithm to solve this problem."}  \\
& Translation : French  \\
& Keyword Removal  : 20\%,40\%,60\%,80\%  \\
& Isotopic Replacement : first verb of each sentence  \\ \midrule
\multirow{2}{*}{\makecell[tc]{Documentation \\ modification}} & Function name masked  \\ 
& Function name + arguments masked  \\ \midrule
\multirow{4}{*}{\makecell[tc]{Documentation \\ modification}} & Shebang  \\ 
& Authors \\
& \textit{- Guido von Rossum}  \\ 
& \textit{- Andrey Petrov}  \\ 
& \textit{- Jean-Baptiste Doderlein}  \\ \midrule
\multirow{3}{*}{\makecell[tc]{Leetcode \\ modification}} & Original \\
& No Documentation  \\
& Function name masked  \\ 
\bottomrule
\end{tabular}
\caption{Variation operators on HumanEval and Leetcode}
\label{table:variation_operator}
\end{table}

To automate the variation of the prompts, we design various variation operators that belong to different categories, such as structural, semantics, and syntax changes. It should be noted that the chosen mutation operators do not necessarily aim at mimicking a realistic developer modification. Some could like the inclusion of generic additional instruction, but others like the removal of keywords are more synthetic. In any case, our primary goal is to assess the sensitivity (or robustness) of Copilot/Codex to prompt variations. 

\subsection{Design of Variation Operators} 
\subsubsection{Documentation modification}
These modifications are applied to the description of the programming task provided in the prompt. 
%For HumanEval and in Python, the docstring of the function provides the documentation on the function to generate.
%
Multiple changes can be applied : 

\begin{itemize}
    \item \textbf{Delete documentation}: Delete the description of the programming task in the prompt.
    \item \textbf{Delete examples}: Remove all examples from the documentation.
    \item \textbf{Additional instruction}: Add some information at the end of the documentation. 
    \item \textbf{Chain of Thought prompt}: Add some word at the end of the documentation and don't close the end of the comment to allow the model to complete the comment before writing the code. 
    \item \textbf{Translation}: Translate the description of the programming task in another language.
    \item \textbf{Keyword Removal}: Remove a certain percentage of the least important words from a programming task description. Words are ranked using the TF-IDF score\cite{Moon2013ASO}, which evaluates the importance of a word in relation to a text corpus.
    \item \textbf{Isotopic Replacement}:Replace some words using the isotopic replacement\cite{sun-improving-2022}. The words are selected in each sentence of the prompt with the model from \cite{iyer-mapping-2018}, a list of possible replacements is generated by BERT Mask\cite{DBLP:journals/corr/abs-1810-04805}, and finally the modified sentence with the highest cosine similarity embedding\cite{reimers-2019-sentence-bert} is chosen for the final prompt.
\end{itemize}

\subsubsection{Signature modification}

The prompt includes the function signature, e.g., its name and arguments.
The tested modification is the hiding of the function name and arguments. A random string replaces the function name and then is replaced by the original function name for testing.

\subsubsection{Context modification}

The context is the part of the prompt that is not specific to the problem. Modifying the context is done by adding additional information at the beginning of the prompt that is not related to the problem itself.

\subsection{Implementation for HumanEval} 
Table \ref{table:variation_operator} shows the different operators we consider in our experiment under the three categories.
We used the variation operators defined above and implemented some variations. 
We tested documentation modification:
\begin{itemize}
    \item Additional instruction with "Write a quick algorithm to solve this problem."
    \item CoT (Chain of Thought) prompt with "Algorithm :", "Complexity :" and "Let's thinks step by step" \cite{ltsbs}
    \item Keyword Removal with 20\%, 40\%, 60\% and 80\%
    \item Isotopic Replacement with the first verb of each sentence with no example in documentation.
\end{itemize}

For signature modification, we have experimented by replacing first just the name of the function by a random string of 8 characters, then by replacing also the arguments. 

Finally, for context modification we used: 
\begin{itemize}
    \item Shebang ("\#!user/bin/env python \# -*- coding: utf-8 -*-") at the beginning of the prompt
    \item Add an author flag at the beginning of the prompt. The authors tested were Guido von Rossum (creator of Python), Andrey Petrov (lead author of Python’s library urllib3) and Jean-Baptiste Doderlein (random name). This modification is inspired by \cite{pearce-asleep-2021}.
\end{itemize}

\subsection{Implementation for Leetcode} 

Only two prompts' variations have been considered on the Leetcode dataset. This choice was made due to the Leetcode API limits (see Section~\ref{subsec:leetcode}) for testing candidate programs generated at different temperatures and over 300+ problems in six programming languages. These two variations were selected based on the results found in the study of the variations of the HumanEval dataset. 
Specifically, we excluded variations that have neutral effect. We retained two prompt variations that question the sensitivity of Codex and Copilot over HumanEval -- more details are given in Section~\ref{subsec:leetcoderes}. %lead to the highest performance drop (worst results). % , and Masked Functions is the one where we got surprising results with HumanEval
 In addition to diversifying the set of problems, considering Leetcode allows one to determine whether results over HumanEval still hold for another dataset.

\section{Empirical Evaluation Results}

% This section presents our results with respect to our research questions. 

\subsection{RQ1: \rqOne} 

To answer this question, we use the original prompts and the default temperature of Copilot and Codex. 
% We recall that pass@k results for Codex and Leetcode are not available due to technical limitations (see Section~\ref{}).
%(1) performance sensitivity to dataset: Leetcode and HumanEval exhibit different characteristics that may impact performance; (2) performance sensitivity to programming language: Leetcode dataset gives the opportunity to try Copilot and Codex over different programming languages; (3) performance evolutiopern over $pass@k$ for Codex over HumanEval. 
Table~\ref{table:rq0datasets} reports the pass@1 results for Codex and Copilot over HumanEval and Leetcode. 
Performance varies between 18.3\% (Leetcode, C language) and 46.3\% (Leetcode, Java). 
Numerous problems are resolved in one shot, but there is room for improvement: Developers cannot trust the generated program in most cases. 
Also, Codex appears to be capable of significantly improving performance when generating more programs (see Table~\ref{table:rq0datasets2}): from 22.44\% at pass@1 to 71.7\% (pass@10) and 98.78\% (closed to perfect, pass@100) on HumanEval and Python3. 

% \emph{Performance sensitivity to dataset.} As early stated in Section~\ref{subsec:datasets}, Leetcode and HumanEval exhibit different characteristics that may impact performance. 
Table~\ref{table:rq0datasets} also shows that, in default settings, Copilot is always superior to Codex, regardless of the dataset and programming language.  
Performance ranking according to the programming languages is highly stable among Copilot and Codex (Spearman correlation: 0.96 with pvalue=0.0004). 
In detail, C is the most difficult programming language, followed by Python while C++ is the most favorable language. 

\begin{tcolorbox}[boxsep=-2pt]
\textbf{$\boldsymbol{RQ_1}$} The performance, though promising, can be improved (46.3\% at best with Java) and varies across programming languages (only 18.3\% for C) for both Copilot and Codex. 
\end{tcolorbox}

\begin{table}\centering
    \ra{1.5}
    
    \begin{tabular}{lcc}
    \toprule
    {} & Copilot & \multicolumn{1}{c}{Codex} \\
    {} & pass@1 & pass@1  \\
    \midrule
    Original (HumanEval, Python3)                      &            31.1 &           22.44  \\ \specialrule{.05em}{0em}{0em}
    % \midrule
    Original (Leetcode, Javascript)  &                    37.3             &                   32.3             \\ \specialrule{.05em}{0em}{0em}
    Original (Leetcode, Java)  &                   46.3              &                 39.3               \\ \specialrule{.05em}{0em}{0em}
 
    Original (Leetcode, C\#)  &                    44.3             &                   33.7             \\ \specialrule{.05em}{0em}{0em}
    Original (Leetcode, Python3)  &                    33.7             &                   30.3             \\ \specialrule{.05em}{0em}{0em}
    %  \midrule
     Original (Leetcode, C)                      &                    18.7             &                  18.3              \\ \specialrule{.05em}{0em}{0em}
     Original (Leetcode, C++)  &                    45.3             &                  41.3              \\ \specialrule{.05em}{0em}{0em}
    \bottomrule
    \end{tabular}

    \caption{Pass@1 (original prompts; default temperature)}
    \label{table:rq0datasets}
\end{table}

\begin{table}\centering
    \ra{1.5}
    \begin{tabular}{lccccc}
   \toprule
    {} & \multicolumn{5}{c}{Codex} \\
    {} & pass@1 &   \red{pass@3} &   \red{pass@5} &  pass@10 &  pass@100 \\
    \midrule
    Original (HumanEval, Python3)  &  22.44 &    44.74   &  56.51 & 71.7 & 98.78 \\ \specialrule{.05em}{0em}{0em}
    \bottomrule
    \end{tabular}
    
    \caption{Pass@k (original prompts; default temperature)}
    \label{table:rq0datasets2}
\end{table}

\subsection{RQ2: \rqTwo}

For answering this question, we vary the original prompt using our operators while keeping the default temperature of Copilot and Codex. 
% For HumanEval, we assess Codex at different pass@k. 
% \emph{Performance sensitivity to programming language.}

\begin{table}\centering
    \ra{1.5}

    \begin{tabular}{lcccccc}
    \toprule
    {} & Copilot & \multicolumn{5}{c}{Codex} \\
    {} & pass@1 & pass@1 &   \red{pass@3} &  \red{pass@5} & pass@10 & pass@100 \\
    \midrule
    Original  &  31.1 & 22.44 & 44.74 & 56.51 & 71.7 & 98.78 \\ \specialrule{.05em}{1em}{0em} 
    No Documentation     &   \textbf{8.54}\textsubscript{\down}\cellcolor{Red!35} &   \textbf{7.74}\textsubscript{\down}\cellcolor{Red!35} & 16.45\textsubscript{\down}\cellcolor{Red!35} & 21.38\textsubscript{\down}\cellcolor{Red!35} & \textbf{28.3}\textsubscript{\down}\cellcolor{Red!35} &  \textbf{64.02}\textsubscript{\down}\cellcolor{Red!35} \\
    No Example & 30.49\cellcolor{Red!15} &  \textbf{17.89}\textsubscript{\down}\cellcolor{Red!35} & 36.35\textsubscript{\down}\cellcolor{Red!35} & 46.11\textsubscript{\down}\cellcolor{Red!35} &\textbf{58.65}\textsubscript{\down}\cellcolor{Red!35} &   \textbf{87.8}\textsubscript{\down}\cellcolor{Red!35} \\
    Algorithm  & 27.44\cellcolor{Red!15} &  \textbf{16.98}\textsubscript{\down}\cellcolor{Red!35} &  35.35\textsubscript{\down}\cellcolor{Red!35} & 45.62\textsubscript{\down}\cellcolor{Red!35} & \textbf{60.09}\textsubscript{\down}\cellcolor{Red!35} &   \textbf{93.9}\textsubscript{\down}\cellcolor{Red!35} \\
    Complexity &  \textbf{16.46}\textsubscript{\down}\cellcolor{Red!35} & 22.85\cellcolor{Green!15} & 45.21\textsubscript{\cellcolor{Green!15}} & 56.54 & 70.69\cellcolor{Red!15} & 96.34\cellcolor{Red!15} \\
    Let's think step by step       &  \textbf{15.85}\textsubscript{\down}\cellcolor{Red!35} &  \textbf{13.49}\textsubscript{\down}\cellcolor{Red!35} &  30.28\textsubscript{\down}\cellcolor{Red!35} & 40.38\textsubscript{\down}\cellcolor{Red!35} & \textbf{54.66}\textsubscript{\down}\cellcolor{Red!35} &  \textbf{89.63}\textsubscript{\down}\cellcolor{Red!35} \\
    Quick     &  \textbf{21.95}\textsubscript{\down}\cellcolor{Red!35} &  \textbf{17.38}\textsubscript{\down}\cellcolor{Red!35} & 38.19\textsubscript{\down}\cellcolor{Red!35} & 50.31\textsubscript{\down}\cellcolor{Red!35} & \textbf{66.7}\textsubscript{\down}\cellcolor{Red!35} & 96.95\cellcolor{Red!15} \\
    French    &  \textbf{26.22}\textsubscript{\down}\cellcolor{Red!35} &  \textbf{16.63}\textsubscript{\down}\cellcolor{Red!35} &  34.67\textsubscript{\down}\cellcolor{Red!35} & 44.77\textsubscript{\down}\cellcolor{Red!35} & \textbf{58.62}\textsubscript{\down}\cellcolor{Red!35} &  \textbf{90.85}\textsubscript{\down}\cellcolor{Red!35} \\
    Keyword Cut 20\%      & 29.27\cellcolor{Red!15} & 21.68\cellcolor{Red!15} & 43.19\textsubscript{\cellcolor{Red!15}} & 54.53\textsubscript{\cellcolor{Red!15}} & \textbf{69.52}\textsubscript{\down}\cellcolor{Red!35} & 98.78 \\
    Keyword Cut 40\%      & 29.27\cellcolor{Red!15} &  \textbf{20.83}\textsubscript{\down}\cellcolor{Red!35} &  40.90\textsubscript{\down}\cellcolor{Red!35} & 51.50\textsubscript{\down}\cellcolor{Red!35} & \textbf{66.07}\textsubscript{\down}\cellcolor{Red!35} & 98.17\cellcolor{Red!15} \\
    Keyword Cut 60\%      & 28.05\cellcolor{Red!15} &  \textbf{19.32}\textsubscript{\down}\cellcolor{Red!35} & 37.12\textsubscript{\down}\cellcolor{Red!35} & 46.51\textsubscript{\down}\cellcolor{Red!35} & \textbf{59.96}\textsubscript{\down}\cellcolor{Red!35} & 97.56\cellcolor{Red!15} \\
    Keyword Cut 80\%      &  \textbf{22.56}\textsubscript{\down}\cellcolor{Red!35} &   \textbf{17.5}\textsubscript{\down}\cellcolor{Red!35} &  33.94\textsubscript{\down}\cellcolor{Red!35} & 42.64\textsubscript{\down}\cellcolor{Red!35} & \textbf{54.96}\textsubscript{\down}\cellcolor{Red!35} &  \textbf{93.29}\textsubscript{\down}\cellcolor{Red!35} \\
    Isotopic Replacement & 26.83\cellcolor{Red!15} &  \textbf{16.24}\textsubscript{\down}\cellcolor{Red!35} &  33.89\textsubscript{\down}\cellcolor{Red!35} & 43.48\textsubscript{\down}\cellcolor{Red!35} & \textbf{55.92}\textsubscript{\down}\cellcolor{Red!35} &   \textbf{87.8}\textsubscript{\down}\cellcolor{Red!35} \\ \specialrule{.05em}{1em}{0em}
    Masked function name & 30.49\cellcolor{Red!15} &  \textbf{19.76}\textsubscript{\down}\cellcolor{Red!35} &  41.11\textsubscript{\down}\cellcolor{Red!35} & 52.75\textsubscript{\down}\cellcolor{Red!35} & \textbf{67.84}\textsubscript{\down}\cellcolor{Red!35} & 98.17\cellcolor{Red!15} \\
    Masked function signature      & 28.05\cellcolor{Red!15} &  \textbf{14.12}\textsubscript{\down}\cellcolor{Red!35} &  32.14\textsubscript{\down}\cellcolor{Red!35} & 43.30\textsubscript{\down}\cellcolor{Red!35} & \textbf{59.29}\textsubscript{\down}\cellcolor{Red!35} & 97.56\cellcolor{Red!15} \\ \specialrule{.05em}{1em}{0em}
    Shebang   &  \textbf{17.07}\textsubscript{\down}\cellcolor{Red!35} & 23.05\cellcolor{Green!15} & 44.12\textsubscript{\cellcolor{Red!15}} & 54.96\textsubscript{\cellcolor{Red!15}} & \textbf{69.15}\textsubscript{\down}\cellcolor{Red!35} &  \textbf{95.73}\textsubscript{\down}\cellcolor{Red!35} \\
    Author Guido von Rossum        &  \textbf{21.95}\textsubscript{\down}\cellcolor{Red!35} &    \textbf{23.68}\textsubscript{\up}\cellcolor{Green!35} & 45.02\textsubscript{\cellcolor{Green!15}} & 55.66\textsubscript{\cellcolor{Red!15}} & \textbf{69.53}\textsubscript{\down}\cellcolor{Red!35} & 97.56\cellcolor{Red!15} \\
    Author Andrey Petrov &  \textbf{20.12}\textsubscript{\down}\cellcolor{Red!35} &    \textbf{24.63}\textsubscript{\up}\cellcolor{Green!35} & 46.09\textsubscript{\cellcolor{Green!15}} & 56.72\textsubscript{\cellcolor{Green!15}} & 70.07\cellcolor{Red!15} &  \textbf{95.73}\textsubscript{\down}\cellcolor{Red!35} \\
    Author Jean-Baptiste Doderlein &  \textbf{17.68}\textsubscript{\down}\cellcolor{Red!35} &    \textbf{24.16}\textsubscript{\up}\cellcolor{Green!35} & 45.44\textsubscript{\cellcolor{Green!15}} & 56.17\textsubscript{\cellcolor{Red!15}} & \textbf{70.13}\textsubscript{\down}\cellcolor{Red!35} & 98.17\cellcolor{Red!15} \\
    \bottomrule
    \end{tabular}
    
    \caption{pass@k results over HumanEval with prompt variations and default, fixed temperature. Bold values have $p<0.05$ for independent T-test. \down is with the less alternative hypothesis and \up with the greater alternative hypothesis}
    \label{table:humaneval_only_variation}
\end{table}

\subsubsection{HumanEval} In Table~\ref{table:humaneval_only_variation}, we present the results obtained for Copilot at pass@1, and for Codex at pass@1, pass@10 and pass@100. Each result is compared to the performance of the original prompt with an independent T-test. In each case, less alternative hypothesis, which tests if the result is significantly worse, and greater alternative hypothesis, which tests if the result is significantly better, are made.

\emph{Quantitative observations.} We observe that prompt changes have a negative impact on the results, most significantly. For Copilot, 10 of the 18 variations tested bring significantly lower results. For Codex, the results are also negative in general. Only 3 variations at pass@1 increase the scores: In this configuration, only adding the author at the beginning of the file with Codex in pass@1 provides better scores. The modification of the context of the file (e.g., Shebang and function signature) impacts strongly Copilot, mostly negatively. While the modification of the function signature does not seem to affect the performance of Copilot, the results for Codex pass@1 and pass10 are not as good with significant decrease.  

\emph{Qualitative analysis.} A purely quantitative analysis does not establish whether performance variations are as expected. Therefore, the nature of the operators of prompt variation, and their supposed impacts, deserve to be discussed in relation to quantitative observations. 
Here are a few remarkable insights:
\begin{itemize}
    \item some variations were not supposed to significantly decrease/increase the performance of Codex/Copilot. For example, it is surprising to observe that the inclusion of Shebang or author name has so much impact (e.g., from 31.1\% to 17.07\% for Copilot pass@1). 
    Cutting words in the prompt also is not impacting much Copilot. 
    \item some prompt variations were expected to decrease the performance. It is the case of "No documentation" and results are in line with this intent. 
    \item some prompt variations are more effective with Copilot or Codex. For example, "Complexity" has a significant negative impact with Copilot, but has a positive impact on Codex. A possible explanation is that Copilot really tries to leverage "Complexity" through code refactoring.  
\end{itemize}
Overall, both Copilot and Codex are sensitive to prompt variations, sometimes in an unexpected way -- inclusion of some contextual information, including of some keywords, or reformulation of the prompt.

\begin{table}\centering
    \ra{1.2}
    \begin{tabular}{@{}rcccccccc@{}}
   \toprule
    &Language & \makecell{No \\ Documentation} & \makecell{Masked\\ function\\ name} \\
    \midrule 
    \multirow{6}{*}{\makecell[cc]{Codex}} 
    &   Javascript &      28.7\textsubscript{\red{-3.6}} &      \textbf{17.3}\textsubscript{\red{-15.0}}  \\
    &   Java       &      44.3\textsubscript{\green{+5.0}} &      \textbf{28.7}\textsubscript{\red{-10.6}}  \\
    &   C\#        &      31.0\textsubscript{\red{-2.7}} &      \textbf{14.0}\textsubscript{\red{-19.7}}  \\
    &   Python3    &      \textbf{41.0}\textsubscript{\green{+10.7}} &      \textbf{16.3}\textsubscript{\red{-14}}  \\
    &   C&      \textbf{11.7}\textsubscript{\red{-6.6}} &       \textbf{9.0}\textsubscript{\red{-9.3}} \\
    &   C++        &      46.0\textsubscript{\green{+4.7}} &      \textbf{26.0}\textsubscript{\red{-15.3}} \\ \midrule
    \multirow{6}{*}{\makecell[cc]{Copilot}}
    &   Javascript &      \textbf{25.3}\textsubscript{\red{-12.0}} &      \textbf{30.0}\textsubscript{\red{-7.3}} \\
    &   Java       &      43.7\textsubscript{\red{-2.6}} &      \textbf{32.3}\textsubscript{\red{-14.0}} \\
    &   C\#        &      44.3    &      \textbf{28.3}\textsubscript{\red{-16.0}} \\
    &   Python3    &      34.3\textsubscript{\green{+0.6}} &      \textbf{19.0}\textsubscript{\red{-14.7}} \\
    &   C&      16.0\textsubscript{\red{-2.7}} &      14.3\textsubscript{\red{-4.4}} \\
    &   C++        &      41.3\textsubscript{\red{-4.0}} &      \textbf{26.7}\textsubscript{\red{-18.6}} \\
    \bottomrule
    \end{tabular}
    \caption{Copilot and Codex($t=1$) performance on LeetCode with prompt variation, default temperature, and $k=1$}
    \label{table:leetcode_variation_only}
\end{table}

\subsubsection{Leetcode}
\label{subsec:leetcoderes}
We evaluated two prompt variations over Leetcode and thus different programming languages -- see Table~\ref{table:leetcode_variation_only}.
% We find that for the original prompt, Copilot performs better than Codex. 
% The impact of the variations
\emph{No Documentation} is very dependent on the programming language. 
On Codex, there is a positive impact for Python, Java and C++, but a negative impact for C\#, Javascript and C. 
Considering the nature of the operator, it is an unexpected result. 
Indeed, removing the problem description is supposed to alter the performance significantly (as in HumanEval). 
Yet, in Leetcode, the degradation is not necessarily significant, and even more surprising, "No documentation" can outperform performance with the original prompt (for Python). 
\emph{Function name masked} leads to a decrease in performance for all languages, for both Copilot and Codex. % The impact of formatting problems following the \cite{nguyen-empirical-2022} model has an important impact for Copilot.
Interestingly, the performance decrease of Copilot was not significant and closed to neutral in HumanEval. On Leetcode, masking the function name has a significant impact. 

% As a side note, the two datasets (LeetCode and HumanEval) bring different results and insights. 
% \emph{Cross-analysis of Leetcode and HumanEval.} 

\begin{tcolorbox}[boxsep=-2pt]
\textbf{$\boldsymbol{RQ_2}$}
Both Copilot and Codex are sensitive to prompt variations, sometimes in an unexpected way -- inclusion of some contextual information, inclusion of some keywords, or reformulation of the prompt can impact the performance significantly. 
% However, depending on the model chosen (Copilot, Codex), the metric (pass@1, pass@10, pass@100) or the programming language, modifying the prompt can lead to better or worse results.
However, the effectiveness of a prompt variation operator does not generalize and depends on the model chosen (Copilot, Codex), the number of generated solutions, or the dataset and programming language. 
For instance, fully removing a problem's documentation significantly decreases performance on HumanEval (as expected), but can surprisingly improve performance on Leetcode. % (significantly for Python and Codex). 
% In HumanEval, removing a problem's documentation significantly decreases the number of problems that have been solved (as expected). However, in Leetcode, similar change increases the number of problems that have been solved. 
\end{tcolorbox}

\subsection{RQ3: \rqThree} 

To answer this question, we vary the temperature of Codex over the original prompt and over different variations of the prompt. 
We recall that Copilot does not give access to modify the temperature, while the Codex API provides a parameter to control this parameter.  
Hence, in the following, we report the performance results of Codex only over HumanEval and LeetCode. 

\begin{figure}
    \centering
    \includegraphics[width=0.8\textwidth]{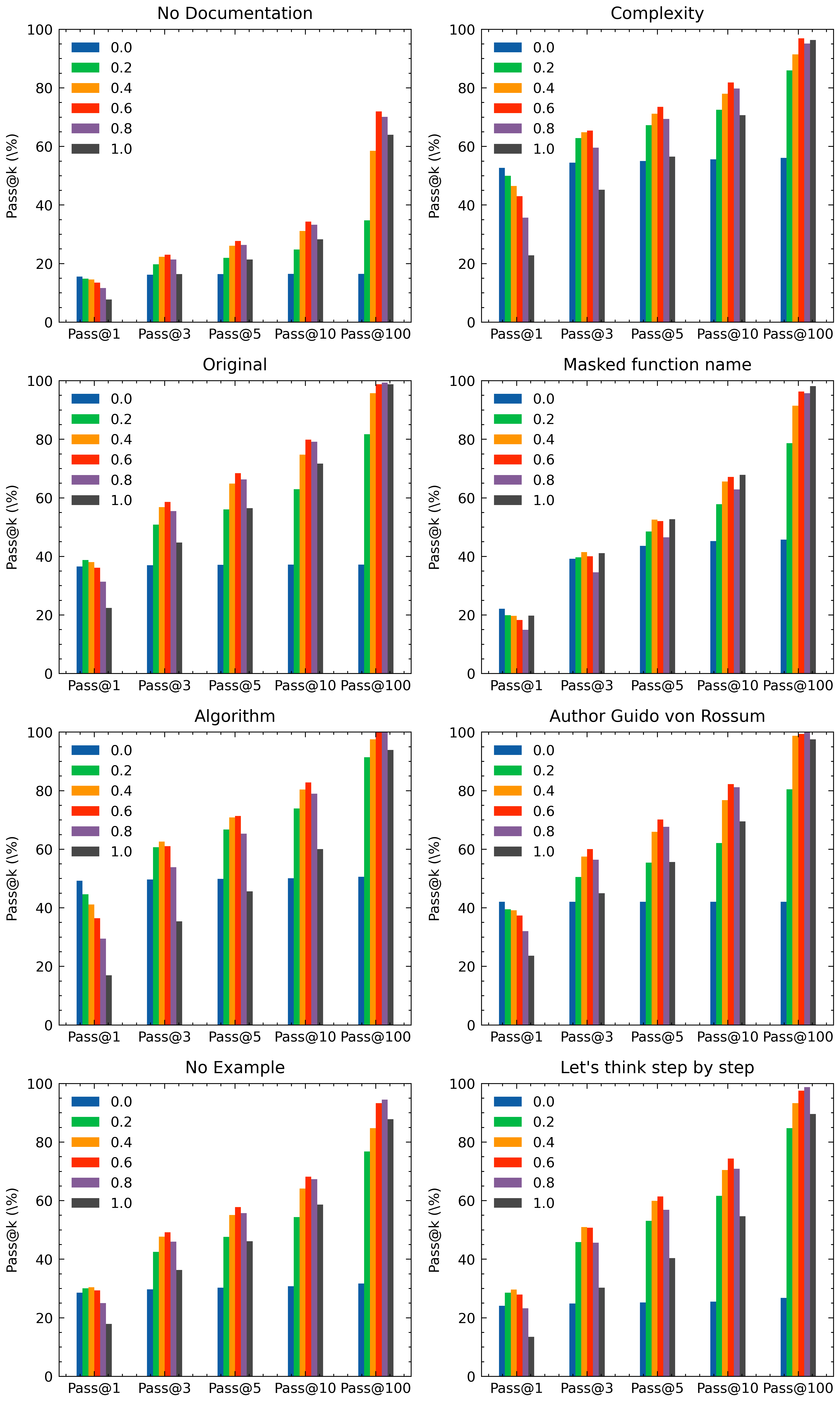}
    \caption{Pass@k Evaluation of Codex on HumanEval dataset for some variation; $temp \in$[0.0,0.2,0.4,0.6,0.8,1]}
    \label{fig:humaneval_passk_codex|variation_temp_k|curated}
\end{figure}

\subsubsection{Codex with varying temperature vs. Copilot, over the original prompt.} 
\begin{figure}
% \vspace*{-4mm}
    \centering
    \includegraphics[width=0.8\textwidth]{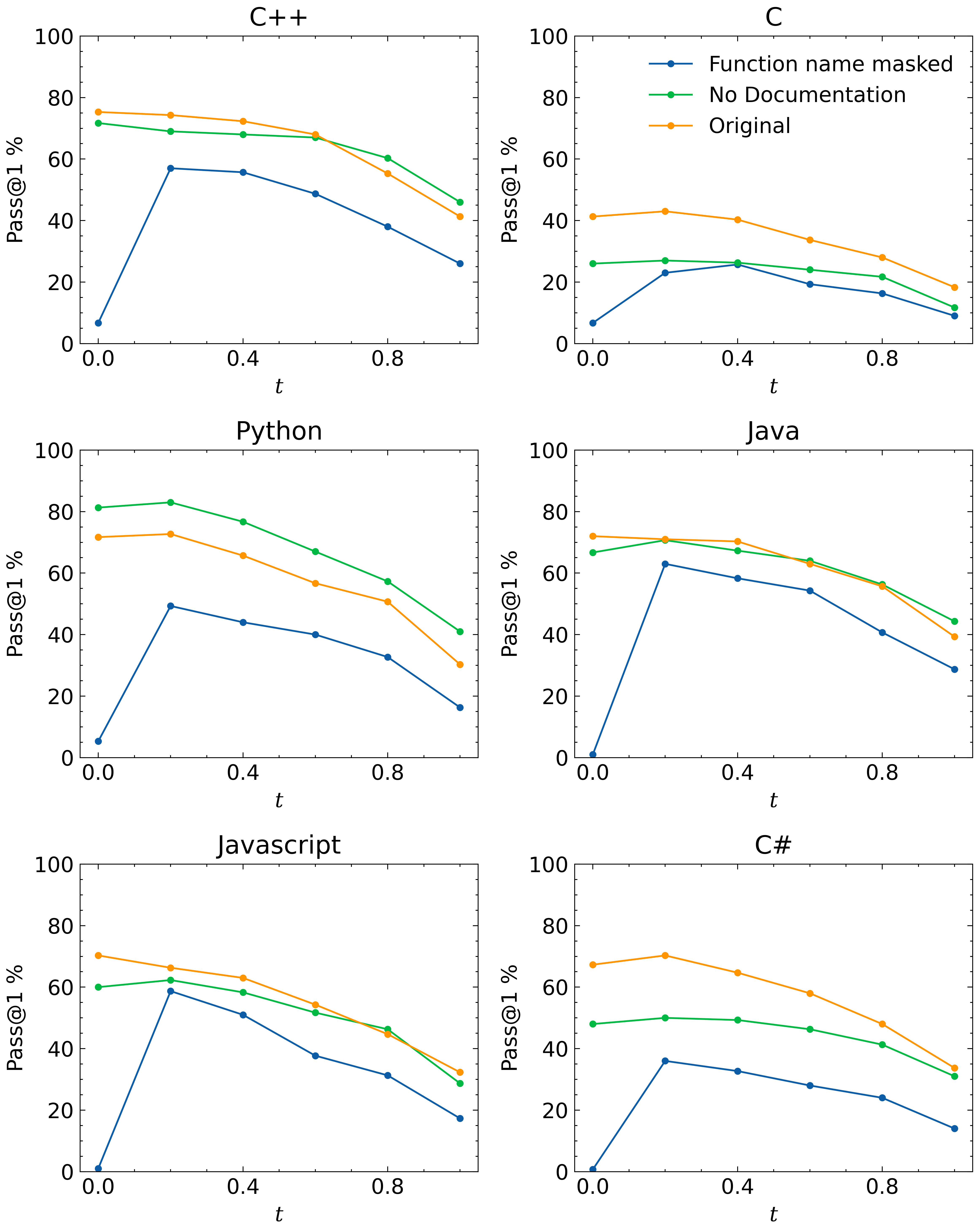} % version noline dispo
    \caption{Pass@1 Evaluation of Codex on Leetcode dataset; $temp \in$[0.0,0.2,0.4,0.6,0.8,1]}
   % \vspace*{-4mm}
    \label{fig:leetcode_pass1_codex}
    % \vspace*{-4mm}
\end{figure}

Figure~\ref{fig:humaneval_passk_codex|variation_temp_k|curated} and Figure \ref{fig:leetcode_pass1_codex} include the performance results of Codex for temperature variations over the original prompt. 
Hence, we can compare with the results of Table~\ref{table:rq0datasets}, RQ1. 
For HumanEval, a temperature $t$ of 0.2 (instead of 1.0 by default) for Codex reaches 38.8, a significant improvement over 22.44. 
Codex with $t=0.2$ also outperforms Copilot at pass@1 (31.1). 
For Leetcode, a temperature of 0.0 or 0.2 is the best for Codex. 
This temperature significantly improves performance for all programming languages (e.g., for JavaScript, Codex $t=0$ reaches 70.3 instead of 32.3 for $t=1$). 
Codex with $t=0.0$ or $t=0.2$ also significantly outperforms Copilot for all languages at pass@1. 
Even non-optimal temperatures of Codex, different from the default one, outperform Copilot for Leetcode and HumanEval. 

\subsubsection{HumanEval} In Figure~\ref{fig:humaneval_passk_codex|variation_temp_k|curated}, we also report the pass@1, \red{pass@3, pass@5,} pass@10 and pass@100 scores for different variations of the prompt at different temperatures. 
A general observation is that there is a complex interplay between temperature, prompt, and number of codes generated. 
For instance, given a prompt variation, the best temperature for the various pass@k differs. If we consider pass@1, we observe two different cases: either the optimal temperature is 0 as in the variations \emph{Complexity}, \emph{No Documentation} and \emph{Masked function name}, or the temperature is around 0.4 for the original prompt and the variations \emph{No Example} and \emph{Let's think step by step}. \blue{For pass@3 and pass@5, the best temperature is around 0.6 in all cases}. For pass@10, the optimal temperature is 0.6, and for pass@100 around 0.6 and 0.8. 
Besides, we observe some particular cases like \emph{Masked function name}, for which the best results for pass@10 and pass@100 are obtained with a temperature of 1.0 (the default temperature of Codex). 
%This can be explained by the need of more creativity  ??? 
That is, some prompt variations are effective only when the temperature is correctly set up with the optimal value. 

% \vspace*{-4mm}

\subsubsection{Leetcode}
Figure \ref{fig:leetcode_pass1_codex} depicts pass@1 of Codex over Leetcode. For each programming language, we report the performance results over temperature variations and three prompts (original and two variations).
A first observation is that the lower the temperature, the better the results for the original prompt. 
\emph{Function name masked} at temperature $t=0$ never exceeds 10\% with an important difference with the original prompt, leading to catastrophic performance.
This prompt variation and specific temperature value should thus be avoided for Leetcode. This contrasts with HumanEval, for which the lower the temperature, the more the result was optimal in pass@1.
Other temperature values combined with \emph{Function name masked} give more adequate results, but the performance is still weaker than with the original prompt.  
% This is, for all languages and temperatures, the most destructive modification. It is always lower than the original.
One would expect to have a similar result with \emph{No Documentation}, which removes the description of the problem leaving only the function signature. 
However, this is not always the case.
On the one hand, for Python, Javascript, Java, and C++, there is at least one temperature where we obtain similar or superior results with \emph{No Documentation} than with the original prompt. This result is particularly visible in Python where we get improvements for all temperatures. 
On the other hand, for C and C\#, \emph{No Documentation} leads to lower performance for all temperatures. 
As in HumanEval, there is a complex interplay between prompt variation and temperature. 
% However, this is not always the case. For C and C\#, \emph{No Documentation} leads to lower performance for all temperatures. However for the other languages, 

\begin{tcolorbox}[boxsep=-2pt]
\textbf{$\boldsymbol{RQ_3}$}
For both HumanEval and LeetCode, varying the temperature can significantly improve Codex performance, up to the point Codex with a configured temperature (not necessarily the best one) can strongly outperform Copilot. 
% There is a complex interplay between temperature, prompt variation, and the number of generated codes
The optimal temperature depends on the chosen prompt variation, the number of generated codes, and to a lesser extent the targeted programming languages. % Some prompt variations become effective under the conditions temperature is correctly set up. 
\end{tcolorbox}

\subsection{RQ4: \rqFour}
% We have seen in the previous results that the choice of the temperature and the prompt variation can have an important impact on pass@k. 
% However, there is a complex interplay: the best temperature varies among prompt variation (and vice-versa). 

% \subsubsection{Best variation and temperature by pass@k}
To answer this research question, we identify what are the best combinations of temperature and prompt variations for Codex that maximize, on average, pass@k over HumanEval.
Table~\ref{table:oracle_single_var_temp_by_pass_k} reports the five best configurations of Codex for pass@1, \red{pass@3, pass@5,} pass@10 and pass@100.  
Each configuration is composed of a prompt variation together with the temperature used. Every result reports the difference with the performance of the default configuration (i.e., the original prompt and temperature $t=1$, see Table~\ref{table:rq0datasets}, RQ1).
At pass@100, the gains are low (+1.22) because the prompt without variation is already very high (98.78). However, we managed to reach 100\%, which means that among the 100 codes generated with these configurations and for each problem, one of the codes passes the test suite.
We get at pass@1 a considerable performance improvement, from 22.44\% to 52.67\%. We observe that the best temperatures for pass@1 are low (0.0, 0.2) contrary to pass@10 or pass@100 for which the best temperatures are closer to 1.0 (0.6, 0.8). 
It is worth noticing that the best configuration for pass@1 is very bad for \blue{pass@5 (-1.47)}, pass@10 (-16.1) and pass@100 (-42.68). 
Intuitively, the prompt variations and temperature used at pass@1 are too conservative at pass@10 and pass@100, limiting Codex to generating creative programs to find a solution.

% It means that prompt variations at pass@10 and pass@100 are not effective for Codex even with the best temperature. 
% Hence, a recommendation is to not vary prompt 

\begin{table}
    \centering
    \ra{1.2}
    \begin{tabular}{@{} r c c c c c c c c c c c c c c c @{}}
    \toprule
    & \multicolumn{2}{c}{Model} & \multicolumn{5}{c}{Score} \\   
    & Variation & $t$ & pass@1 & \red{pass@3} & \red{pass@5} & pass@10 & pass@100  \\ \midrule
    % Reference & Original & 1.0 & 22.44 & 71.70 & 98.78 \\ \midrule
    %pass 100
    \multirow{5}{*}{\makecell[t c]{Best \\ pass@100}}
    & Author Ap   	&	0.6	&	38.15\textsubscript{\green{+15.71}}	&	60.12\textsubscript{\green{+15.38}}	&	69.57\textsubscript{\green{+13.06}}	&	80.53\textsubscript{\green{+8.83}} 	&	100.0 \textsubscript{\green{+1.22}}	\\
    & Algorithm   	&	0.6	&	36.44\textsubscript{\green{+14.0}}	&	61.04\textsubscript{\green{+16.3}}	&	71.4\textsubscript{\green{+14.89}}	&	82.84\textsubscript{\green{+11.14}} 	&	100.0 \textsubscript{\green{+1.22}}	\\
    &     	&	0.8	&	29.46\textsubscript{\green{+7.02}}	&	53.9\textsubscript{\green{+9.16}}		&	65.32\textsubscript{\green{+8.81}}	&	78.99\textsubscript{\green{+7.29}} 	&	100.0 \textsubscript{\green{+1.22}}	\\
    & Author Gvr   	&	0.8	&	32.07\textsubscript{\green{+9.63}}	&	56.44\textsubscript{\green{+11.7}}	&	67.66\textsubscript{\green{+11.15}}	&	81.2\textsubscript{\green{+9.5}} 	&	100.0 \textsubscript{\green{+1.22}}	\\
    &     	&	0.6	&	37.37\textsubscript{\green{+14.93}}	&	60.03\textsubscript{\green{+15.29}}	&	70.16\textsubscript{\green{+13.65}}	&	82.25\textsubscript{\green{+10.55}} 	&	99.39 \textsubscript{\green{+0.61}}	\\
    \midrule
    
    %pass 10
    \multirow{5}{*}{\makecell[t c]{Best \\ pass@10}}
    & Algorithm   	&	0.6	&	36.44\textsubscript{\green{+14.0}}	&	61.04\textsubscript{\green{+16.3}}	&	71.4\textsubscript{\green{+14.89}}	&	82.84 \textsubscript{\green{+11.14}}	&	100.0\textsubscript{\green{+1.22}} \\
    & Author Gvr   	&	0.6	&	37.37\textsubscript{\green{+14.93}}	&	60.03\textsubscript{\green{+15.29}}	&	70.16\textsubscript{\green{+13.65}}	&	82.25 \textsubscript{\green{+10.55}}	&	99.39\textsubscript{\green{+0.61}} \\
    & Complexity   	&	0.6	&	43.02\textsubscript{\green{+20.58}}	&	65.42\textsubscript{\green{+20.68}}	&	73.56\textsubscript{\green{+17.05}}	&	81.85 \textsubscript{\green{+10.15}}	&	96.95\textsubscript{\red{-1.83}} \\
    & Shebang   	&	0.6	&	38.24\textsubscript{\green{+15.8}}	&	60.39\textsubscript{\green{+15.65}}	&	69.99\textsubscript{\green{+13.48}}	&	81.41 \textsubscript{\green{+9.71}}		&	99.39\textsubscript{\green{+0.61}} \\
    & Author Gvr   	&	0.8	&	32.07\textsubscript{\green{+9.63}}	&	56.44\textsubscript{\green{+11.7}}	&	67.66\textsubscript{\green{+11.15}}	&	81.2 \textsubscript{\green{+9.5}}		&	100.0\textsubscript{\green{+1.22}} \\
    \midrule
    
    %pass 5
    \multirow{5}{*}{\makecell[t c]{\red{Best} \\ \red{pass@5}}}
    & Complexity   	&	0.6	&	43.02\textsubscript{\green{+20.58}}	&	65.42\textsubscript{\green{+20.68}}	&	73.56 \textsubscript{\green{+17.05}}	&	81.85\textsubscript{\green{+10.15}}	&	96.95\textsubscript{\red{-1.83}} \\
    & Algorithm   	&	0.6	&	36.44\textsubscript{\green{+14.0}}	&	61.04\textsubscript{\green{+16.3}}	&	71.4 \textsubscript{\green{+14.89}}		&	82.84\textsubscript{\green{+11.14}}	&	100.0\textsubscript{\green{+1.22}} \\
    & Complexity   	&	0.4	&	46.5\textsubscript{\green{+24.06}}	&	64.89\textsubscript{\green{+20.15}}	&	71.2 \textsubscript{\green{+14.69}}		&	78.03\textsubscript{\green{+6.33}}	&	91.46\textsubscript{\red{-7.32}} \\
    & Algorithm   	&	0.4	&	41.18\textsubscript{\green{+18.74}}	&	62.62\textsubscript{\green{+17.88}}	&	70.87 \textsubscript{\green{+14.36}}	&	80.43\textsubscript{\green{+8.73}}	&	97.56\textsubscript{\red{-1.22}} \\
    & Author Gvr   	&	0.6	&	37.37\textsubscript{\green{+14.93}}	&	60.03\textsubscript{\green{+15.29}}	&	70.16 \textsubscript{\green{+13.65}}	&	82.25\textsubscript{\green{+10.55}}	&	99.39\textsubscript{\green{+0.61}} \\
    \midrule
    
    %pass 3
    \multirow{5}{*}{\makecell[t c]{\red{Best} \\ \red{pass@3}}}
    & Complexity   	&	0.6	&	43.02\textsubscript{\green{+20.58}}	&	65.42 \textsubscript{\green{+20.68}}	&	73.56\textsubscript{\green{+17.05}}	&	81.85\textsubscript{\green{+10.15}}	&	96.95\textsubscript{\red{-1.83}} \\
    &     	&	0.4	&	46.5\textsubscript{\green{+24.06}}	&	64.89 \textsubscript{\green{+20.15}}	&	71.2\textsubscript{\green{+14.69}}	&	78.03\textsubscript{\green{+6.33}}	&	91.46\textsubscript{\red{-7.32}} \\
    &     	&	0.2	&	49.99\textsubscript{\green{+27.55}}	&	62.85 \textsubscript{\green{+18.11}}	&	67.26\textsubscript{\green{+10.75}}	&	72.51\textsubscript{\green{+0.81}}	&	85.98\textsubscript{\red{-12.8}} \\
    & Algorithm   	&	0.4	&	41.18\textsubscript{\green{+18.74}}	&	62.62 \textsubscript{\green{+17.88}}	&	70.87\textsubscript{\green{+14.36}}	&	80.43\textsubscript{\green{+8.73}}	&	97.56\textsubscript{\red{-1.22}} \\
    &     	&	0.6	&	36.44\textsubscript{\green{+14.0}}	&	61.04 \textsubscript{\green{+16.3}}		&	71.4\textsubscript{\green{+14.89}}	&	82.84\textsubscript{\green{+11.14}}	&	100.0\textsubscript{\green{+1.22}} \\
    \midrule
    
    %pass 1
    \multirow{5}{*}{\makecell[t c]{Best \\ pass@1}}
    & Complexity   	&	0.0	&	52.67 \textsubscript{\green{+30.23}}	&	54.46\textsubscript{\green{+9.72}}	&	55.04\textsubscript{\red{-1.47}}	&	55.6\textsubscript{\red{-16.1}}	&	56.1\textsubscript{\red{-42.68}} \\
    &     	&	0.2	&	49.99 \textsubscript{\green{+27.55}}	&	62.85\textsubscript{\green{+18.11}}	&	67.26\textsubscript{\green{+10.75}}	&	72.51\textsubscript{\green{+0.81}}	&	85.98\textsubscript{\red{-12.8}} \\
    & Algorithm   	&	0.0	&	49.22 \textsubscript{\green{+26.78}}	&	49.7\textsubscript{\green{+4.96}}	&	49.92\textsubscript{\red{-6.59}}	&	50.1\textsubscript{\red{-21.6}}	&	50.61\textsubscript{\red{-48.17}} \\
    & Complexity   	&	0.4	&	46.5 \textsubscript{\green{+24.06}}		&	64.89\textsubscript{\green{+20.15}}	&	71.2\textsubscript{\green{+14.69}}	&	78.03\textsubscript{\green{+6.33}}	&	91.46\textsubscript{\red{-7.32}} \\
    & Algorithm   	&	0.2	&	44.62 \textsubscript{\green{+22.18}}	&	60.73\textsubscript{\green{+15.99}}	&	66.76\textsubscript{\green{+10.25}}	&	73.95\textsubscript{\green{+2.25}}	&	91.46\textsubscript{\red{-7.32}} \\
    \bottomrule
    \end{tabular}
    \caption{Best prompt variations and temperatures for Codex on HumanEval at pass@1, \red{pass@3, pass@5,} pass@10 and pass@100. \blue{The \textit{Model Variation} column ranks the top five variations with their associated temperatures for each pass@100. For example, under pass@100, the best configuration is\textit{Author Ap} with a temperature of 0.6, achieving a score of 100, which is an improvement of +1.22 compared to the Original variation for the same pass@k. The other values on the same line represent the scores for different pass@k values. For instance, under pass@3, the same configuration (Author Ap at temperature 0.6) scores 60.12, showing a gain of 15.38 over the Original variation at pass@3.}}
    \label{table:oracle_single_var_temp_by_pass_k}
\end{table}

\begin{tcolorbox}[boxsep=-2pt]
\textbf{$\boldsymbol{RQ_4}$}
A proper configuration of the temperature and prompt variation can significantly improve the results of Codex. 
% An even better strategy is to combine different configurations when generating $k > 1$ programs. 
 A difficulty, though, is that the best configurations differ for a given pass@k.
A general recommendation is to use low temperatures (0.0, 0.2) for pass@1, and higher temperatures (0.6 to 1.0) for pass@10 and @100. 
% Moreover, the best configurations for pass@1 or pass@2 produce very bad results for pass@10 or pass@100
\end{tcolorbox}

\subsection{RQ5: \rqFive}

For answering this question, we retrospectively look at each individual problem of HumanEval and report on the best performance that can be achieved when modifying the prompt and the temperature of Codex.  
% We suppose in this section the possibility to have an oracle that for each problem would be able to give us the same number of parameters. 
 Even if unrealistic, the results are interesting to evaluate the potential of a tuning strategy. 

\subsubsection{Fixed Temperature}
\begin{table}\centering
    \ra{1.2}
    \begin{tabular}{lrrrrr}
\toprule
Temperature &  pass@1 & \red{pass@3} &  \red{pass@5} & pass@10 &  pass@100 \\
\midrule
Original $t=1.0$ & 22.44 & 44.74 & 56.51 & 71.70 & 98.78 \\ \midrule
0.0 & 74.99\textsubscript{\green{+52.55}} & 77.76\textsubscript{\green{+33.02}} & 78.58\textsubscript{\green{+22.07}} & 79.06\textsubscript{\green{+7.36}} & 79.27\textsubscript{\red{-19.51}} \\
0.2 & 65.23\textsubscript{\green{+42.79}} & 79.75\textsubscript{\green{+35.01}} & 84.87\textsubscript{\green{+28.36}} & 90.43\textsubscript{\green{+18.73}} & 99.39\textsubscript{\green{+0.61}} \\
0.4 & 56.83\textsubscript{\green{+34.39}} & 77.44\textsubscript{\green{+32.70}} & 84.85\textsubscript{\green{+28.34}} & 92.5\textsubscript{\green{+20.80}} & 100.0\textsubscript{\green{+1.22}} \\
0.6 & 49.93\textsubscript{\green{+27.49}} & 74.49\textsubscript{\green{+29.75}} & 83.61\textsubscript{\green{+27.10}} & 92.49\textsubscript{\green{+20.79}} & 100.0\textsubscript{\green{+1.22}} \\
0.8 & 41.37\textsubscript{\green{+18.93}} & 67.93\textsubscript{\green{+23.19}} & 78.9\textsubscript{\green{+22.39}} & 90.45\textsubscript{\green{+18.75}} & 100.0\textsubscript{\green{+1.22}} \\
1.0 & 30.47\textsubscript{\green{+8.03}} & 55.99\textsubscript{\green{+11.25}} & 68.24\textsubscript{\green{+11.73}} & 83.01\textsubscript{\green{+11.31}} & 100.0\textsubscript{\green{+1.22}} \\
\bottomrule
\end{tabular}
    \caption{Pass@k results for each temperature of Codex, with the best variation for each problem}
    \label{fig:oracle_fixed_temperature}
\end{table}
First, we are interested in identifying the best temperature values. To do so, we fix a temperature value, select the best prompt variation, and report on pass@1, pass@10, and pass@100 for each problem. 
Table~\ref{fig:oracle_fixed_temperature} depicts the results. A first observation is that we can significantly improve the performance of the default configuration (up to 74.99\% at pass@1 for $t=0$). 
A tuning per problem also significantly outperforms the best \emph{average} configurations (see RQ4, Table~\ref{table:oracle_single_var_temp_by_pass_k}). 
Hence, it is worth combining temperature and prompt variation. 
However, the default temperature ($t=1$) of Codex is by far the less effective at pass@1 and should be avoided in our context, as it is poorly effective regardless of the prompt variation. 
Also, the general recommendation is to use a temperature of 0.0 for pass@1, 0.4 for pass@10, and from 0.4 to 1.0 for pass@100. 

% for each problem, the result of the best prompt variation from Table~\ref{table:variation_operator} is used 
% \todo{can we add the variation, cuz we don't see it, only temp -> -> hidden because different for each of 164 problems}

\begin{table}\centering
    \ra{1.2}
    \begin{tabular}{lrrrrr}
\toprule
Variation &  pass@1 & \red{pass@3} & \red{pass@5} & pass@10 &  pass@100 \\
\midrule
Original $t=1.0$ & 22.44 & 44.74 & 56.51 & 71.70 & 98.78 \\ \midrule
Original & 50.15\textsubscript{\green{+27.71}} & 65.69\textsubscript{\green{+20.95}} & 74.13\textsubscript{\green{+17.62}} & 84.81\textsubscript{\green{+13.11}} & 100.0\textsubscript{\green{+1.22}} \\
No Documentation & 20.23\textsubscript{\red{-2.21}} & 27.5\textsubscript{\red{-17.24}} & 32.22\textsubscript{\red{-24.29}} & 40.08\textsubscript{\red{-31.62}} & 89.63\textsubscript{\red{-9.15}} \\
No Example & 41.82\textsubscript{\green{+19.38}} & 56.22\textsubscript{\green{+11.48}} & 63.84\textsubscript{\green{+7.33}} & 73.94\textsubscript{\green{+2.24}} & 98.78\textsubscript{\green{+0.00}} \\
Algorithm & 59.13\textsubscript{\green{+36.69}} & 71.42\textsubscript{\green{+26.68}} & 78.39\textsubscript{\green{+21.88}} & 87.29\textsubscript{\green{+15.59}} & 100.0\textsubscript{\green{+1.22}} \\
Complexity & 62.98\textsubscript{\green{+40.54}} & 74.64\textsubscript{\green{+29.90}} & 80.07\textsubscript{\green{+23.56}} & 86.63\textsubscript{\green{+14.93}} & 98.17\textsubscript{\red{-0.61}} \\
Let’s think step by step & 43.07\textsubscript{\green{+20.63}} & 60.54\textsubscript{\green{+15.80}} & 69.22\textsubscript{\green{+12.71}} & 80.32\textsubscript{\green{+8.62}} & 100.0\textsubscript{\green{+1.22}} \\
Quick & 40.9\textsubscript{\green{+18.46}} & 56.54\textsubscript{\green{+11.80}} & 66.15\textsubscript{\green{+9.64}} & 79.36\textsubscript{\green{+7.66}} & 99.39\textsubscript{\green{+0.61}} \\
French & 38.85\textsubscript{\green{+16.41}} & 52.97\textsubscript{\green{+8.23}} & 61.23\textsubscript{\green{+4.72}} & 72.7\textsubscript{\green{+1.00}} & 98.78\textsubscript{\green{+0.00}} \\
Isotopic Replacement & 41.07\textsubscript{\green{+18.63}} & 54.77\textsubscript{\green{+10.03}} & 62.7\textsubscript{\green{+6.19}} & 73.74\textsubscript{\green{+2.04}} & 98.17\textsubscript{\red{-0.61}} \\
Masked function name & 29.42\textsubscript{\green{+6.98}} & 55.42\textsubscript{\green{+10.68}} & 66.19\textsubscript{\green{+9.68}} & 78.09\textsubscript{\green{+6.39}} & 99.39\textsubscript{\green{+0.61}} \\
Masked function signature & 47.84\textsubscript{\green{+25.40}} & 61.63\textsubscript{\green{+16.89}} & 69.92\textsubscript{\green{+13.41}} & 81.42\textsubscript{\green{+9.72}} & 100.0\textsubscript{\green{+1.22}} \\
Keyword Cut 20\% & 47.27\textsubscript{\green{+24.83}} & 62.9\textsubscript{\green{+18.16}} & 71.99\textsubscript{\green{+15.48}} & 84.14\textsubscript{\green{+12.44}} & 100.0\textsubscript{\green{+1.22}} \\
Keyword Cut 40\% & 46.81\textsubscript{\green{+24.37}} & 60.81\textsubscript{\green{+16.07}} & 69.38\textsubscript{\green{+12.87}} & 81.53\textsubscript{\green{+9.83}} & 100.0\textsubscript{\green{+1.22}} \\
Keyword Cut 60\% & 44.21\textsubscript{\green{+21.77}} & 57.22\textsubscript{\green{+12.48}} & 65.4\textsubscript{\green{+8.89}} & 77.01\textsubscript{\green{+5.31}} & 98.78\textsubscript{\green{+0.00}} \\
Keyword Cut 80\% & 40.03\textsubscript{\green{+17.59}} & 51.54\textsubscript{\green{+6.80}} & 59.02\textsubscript{\green{+2.51}} & 70.87\textsubscript{\red{-0.83}} & 98.78\textsubscript{\green{+0.00}} \\
Shebang & 54.39\textsubscript{\green{+31.95}} & 67.85\textsubscript{\green{+23.11}} & 75.45\textsubscript{\green{+18.94}} & 85.59\textsubscript{\green{+13.89}} & 100.0\textsubscript{\green{+1.22}} \\
Author Guido von Rossum & 53.96\textsubscript{\green{+31.52}} & 68.24\textsubscript{\green{+23.50}} & 76.52\textsubscript{\green{+20.01}} & 87.12\textsubscript{\green{+15.42}} & 100.0\textsubscript{\green{+1.22}} \\
Author Andrey Petrov & 54.82\textsubscript{\green{+32.38}} & 68.97\textsubscript{\green{+24.23}} & 76.71\textsubscript{\green{+20.20}} & 86.19\textsubscript{\green{+14.49}} & 100.0\textsubscript{\green{+1.22}} \\
Author Jean-Baptiste Doderlein & 52.93\textsubscript{\green{+30.49}} & 66.91\textsubscript{\green{+22.17}} & 74.7\textsubscript{\green{+18.19}} & 84.76\textsubscript{\green{+13.06}} & 100.0\textsubscript{\green{+1.22}} \\
\bottomrule
\end{tabular}
    \caption{Pass@k results for each variation, with the best temperature for each problem}
    \label{fig:oracle_fixed_variation}
\end{table}
%Commentaire : 
\subsubsection{Fixed Prompt}
We are interested in identifying the best prompt variations. We fix the variation and take the best temperature($t \in [0.0,0.2,0.4,0.6,0.8,1.0]$) for each problem. 
Table~\ref{fig:oracle_fixed_variation} reports on the results. There are several prompt variations that can significantly outperform the default configuration at pass@1 and pass@10. 
The best variation is \emph{Complexity} at pass@1, but even cutting words strategies are effective. 
\emph{No documentation} on HumanEval has strong limitations at pass@1, \red{pass@3, pass@5,} pass@10, and pass@100, even with the best temperature.  
In general, results confirm the crucial role of the temperature for fully realizing the potential of prompt variation operators.
% at pass@10 \emph{Author Guido von Rossum}, and many variations get 100\% at pass@100 including the original prompt.

\begin{table}\centering
    \ra{1.2}
    
\begin{tabular}{llllll}
\toprule
{} & pass@1 & \red{pass@3} & \red{pass@5} & pass@10 &        pass@100 \\
\midrule
Original $t=1.0$ & 22.44 & 44.74 & 56.51 & 71.70 & 98.78 \\
Oracle & 79.27\textsubscript{\green{+56.83}} & 87.24\textsubscript{\green{+42.50}} & 91.13\textsubscript{\green{+34.62}} & 95.59\textsubscript{\green{+23.89}} & 100.0\textsubscript{\green{+1.22}} \\
\bottomrule
\end{tabular}

    \caption{Pass@k results for best overall, with the best temperature and variation for each problem}
    \label{fig:oracle_best_overall}
\end{table}
\subsubsection{Best overall}

Finally, we consider the case where we took the best temperature and the best variation for each problem. We obtain 79.27\% at pass@1, 95.59\% at pass@10 and 100\% at pass@100 (see "Oracle" in Table~\ref{fig:oracle_best_overall}). 
It is a great improvement over the default configuration of Codex, but also over the best average configurations (RQ4). It is also worth tuning temperature and prompt \emph{together}. 
This result is unrealistic and unusable in practice since there is no \emph{a priori} strategy to determine the best combination of temperature and prompt for each problem. 
However, it gives an upper bound to the achievable results with the variation of these parameters. %\todo{can we give the three best combination ? no because this is the best result for every parameter}
It calls to investigate in the future a \emph{per-problem auto-tuning} capable of selecting the best temperature and/or prompt variations. 

\begin{tcolorbox}[boxsep=-2pt]
\textbf{$\boldsymbol{RQ_5}$} Tuning temperature and prompt variation per problem can provide strong performance improvements: up to 79.27\% at pass@1 compared to 22.44\% for the default configuration and outperforming the best average configuration of RQ4 (52.67\%). 
% The temperature is the most important factor, being both an enactor for prompt variation 
\end{tcolorbox}

%% replication main text

\red{
\section{Study Reproduction with StarCoder2-15B on HumanEval}

\red{With the new release of various LLMs, we are tempted to replicate our original experiment from back in 2022 with Codex language models and Copilot. 
We select the \emph{fully} open source (not only open weights) StarCoder2-15B~\cite{starcoder2} a 15B parameter model released in february 2024 and trained on 600+ programming languages from The Stack v2 public dataset.} 

%After experimenting with Codex language models and Copilot on the HumanEval and LeetCode benchmarks, we achieve the same experience to validate the hypothesis of prompt sensitivity with StarCoder2-15B over HumanEval benchmark only.
%StarCoder2-15B model is a 15B parameter model released in february 2024, trained on 600+ programming languages from The Stack v2 dataset. \\

The reproduction follows the same research questions as above:
\newcommand{\rqSixOne}[0]{What is the performance of StarCoder2-15B over HumanEval?} 
\newcommand{\rqSixTwo}[0]{What is the impact of prompt variation on StarCoder2-15B performance?}
\newcommand{\rqSixThree}[0]{What is the impact of the variation of both temperature and prompts on StarCoder2-15B performance?}
\newcommand{\rqSixFour}[0]{What are the best prompt variations and temperatures for StarCoder2-15B to maximize its performance over a given set of problems?}
\newcommand{\rqSixFive}[0]{How much can we improve the performance of StarCoder2-15B by tuning the temperature and prompts \emph{per} a specific problem?}

\begin{itemize}
    \item[RQ6.1] \emph{\rqSixOne}~
    We use the original prompts and the default temperature (0.2) of StarCoder2-15B. 
At this step, we do not make vary the prompts or the temperature. 
Our goal is to establish a first trend and a baseline for the performance of StarCoder2-15B and the 2 previous models. We also aim to gather insights about performance sensitivity over HumanEval dataset.
    \item[RQ6.2] \emph{\rqSixTwo~}
    Our goal is to assess the sensitivity of StarCoder2-15B with respect to prompt modification.  
    We make vary the prompts, leveraging different operators acting over the programming task description, and some information from the surrounding context (e.g., function signature). However, we rely on default temperature of StarCoder2-15B.
    \item[RQ6.3] \emph{\rqSixThree~}
    We aim to establish the impact of varying all input parameters, including temperature, prompt, and number $k$ of generated solutions.
    \item[RQ6.4] \emph{\rqSixFour~}
    We aim to identify the best parameters for StarCoder2-15B to maximize its performance for a given set of problems (on average). 
    \item[RQ6.5] \emph{\rqSixFive~}
    We aim to identify the best parameters for StarCoder2-15B to maximize its performance for a given problem. Intuitively, the best parameters on average are not necessarily the best ones for a specific problem. 
\end{itemize}

\subsection{Empirical Evaluation Results}

\red{This section presents our replication results with respect to our research questions. 
Note that due to the scalability needs for Leetcode and its cost, we replicated our original study only with the HumanEval dataset.} 

\subsubsection{RQ6.1: \rqSixOne} 

\begin{table}\centering
    \ra{1.5}
    \begin{tabular}{lccccc}
   \toprule
    {} & \multicolumn{5}{c}{StarCoder2-15B} \\
    {} & pass@1 &    pass@3 & pass@5 &     pass@10 &        pass@100 \\
    \midrule
    Original $t=0.2$ (HumanEval, Python3)  &  32.74 & 40.50 & 44.06 & 48.88 & 63.41 \\ \specialrule{.05em}{0em}{0em}
    \bottomrule
    \end{tabular}
    
    \caption{Pass@k (original prompts; default temperature)}
    \label{table:rq0datasets_61}
\end{table}

To answer this question, we report the results on Table~\ref{table:rq0datasets_61}. With the default temperature (0.2) of StarCoder2-15B. Running StarCoder2-15B with same configurations as in Anton et al.~\cite{lozhkov2024starcoder} (official paper), their pass@1 score (46.3\%) is higher than ours (32.74\%), a difference of 13.56\%.
Also, StarCoder2-15B appears to be capable of significantly improving performance when generating more programs (cf. Table~\ref{table:rq0datasets_61}): from 32.74\% at pass@1 to 48.88\% (pass@10) and 63.41\% at pass@100. 

\begin{tcolorbox}[boxsep=-2pt]
\textbf{$\boldsymbol{RQ6.1}$} As in the original study, we observe that performance can be improved by generating many solutions for the same problem. \red{However, while StarCoder2-15B performs better at pass@1 (32.74\% vs 22.44\%), it performs worst at pass@10 (48.88\% vs 71.7\%) and pass@100 (63.41\% vs 98.78\%) in comparison to Codex with default configurations.} \blue{Also, pass@3 and pass@5 scores are better than pass@1.}
\end{tcolorbox}

\subsubsection{RQ6.2: \rqSixTwo} 
For answering this question, we also varied the original prompt using the same operators and kept the default temperature (0.2). 
In Table~\ref{table:humaneval_only_variation_sc}, we present the results obtained at pass@1, pass@3, pass@5, pass@10 and pass@100. Each result is compared to the performance of the original prompt. We observe that prompt change have negative impact on the results generally. The effectiveness of a prompt variation operator does not generalize and depends of the number of generated solutions.

\begin{tcolorbox}[boxsep=-2pt]
\textbf{$\boldsymbol{RQ6.2}$} While Codex and Copilot is negatively impacted by the prompt variation in all cases except $Complexity$ and all authors modifications, with StarCoder2-15B, \red{we observe the same behavior. However, we also observe that the $Complexity$ variation significantly improved the results in all passes.} 
\end{tcolorbox}

\begin{table}\centering
    \ra{1.5}
    \begin{tabular}{lccccc}
   \toprule
    {} & \multicolumn{5}{c}{StarCoder2-15B} \\
    {} & pass@1 & pass@3 & pass@5 &  pass@10 &  pass@100 \\
    \midrule
    Original $t=0.2$ (HumanEval, Python3)& 32.74 & 40.50 & 44.06 & 48.88 & 63.41 \\ \specialrule{.05em}{0em}{0em}
    No Documentation & 9.08\textsubscript{\down}\cellcolor{Red!35} & 13.13\textsubscript{\down}\cellcolor{Red!35} & 14.65\textsubscript{\down}\cellcolor{Red!35} & 16.51\textsubscript{\down}\cellcolor{Red!35} & 22.56\textsubscript{\down}\cellcolor{Red!35} \\
    No Example & 29.84\textsubscript{\down}\cellcolor{Red!35} & 37.86\textsubscript{\down}\cellcolor{Red!35} & 41.54\textsubscript{\down}\cellcolor{Red!35} & 46.58\textsubscript{\cellcolor{Red!15}} & 59.76\textsubscript{\down}\cellcolor{Red!35} \\
    Algorithm & 40.87\textsubscript{\up}\cellcolor{Green!35} & 55.62\textsubscript{\up}\cellcolor{Green!35} & 61\textsubscript{\up}\cellcolor{Green!35} & 67.33\textsubscript{\up}\cellcolor{Green!35} & 81.71\textsubscript{\up}\cellcolor{Green!35} \\
    Complexity & 44.87\textsubscript{\up}\cellcolor{Green!35} & 55.87\textsubscript{\up}\cellcolor{Green!35} & 60.08\textsubscript{\up}\cellcolor{Green!35} & 65.59\textsubscript{\up}\cellcolor{Green!35} & 78.05\textsubscript{\up}\cellcolor{Green!35} \\
    Let’s think step by step & 22.84\textsubscript{\down}\cellcolor{Red!35} & 36.78\textsubscript{\down}\cellcolor{Red!35} & 42.88\textsubscript{\cellcolor{Red!15}} & 50.51\textsubscript{\cellcolor{Green!15}} & 71.95\textsubscript{\up}\cellcolor{Green!35} \\
    Quick & 24.61\textsubscript{\down}\cellcolor{Red!35} & 32.54\textsubscript{\down}\cellcolor{Red!35} & 35.90\textsubscript{\down}\cellcolor{Red!35} & 40.14\textsubscript{\down}\cellcolor{Red!35} & 55.49\textsubscript{\down}\cellcolor{Red!35} \\
    French & 26.08\textsubscript{\down}\cellcolor{Red!35} & 32.79\textsubscript{\down}\cellcolor{Red!35} & 35.87\textsubscript{\down}\cellcolor{Red!35} &  40.35\textsubscript{\down}\cellcolor{Red!35} & 56.71\textsubscript{\down}\cellcolor{Red!35} \\
    Keyword Cut 20\% & 34.61\textsubscript{\up}\cellcolor{Green!35} & 42.28\textsubscript{\cellcolor{Green!15}} & 45.20\textsubscript{\cellcolor{Green!15}} & 49.21\cellcolor{Green!15} & 60.98\textsubscript{\down}\cellcolor{Red!35} \\
    Keyword Cut 40\% & 33.43\cellcolor{Green!15} & 40.26\cellcolor{Red!15} & 42.89\cellcolor{Red!15} & 46.37\textsubscript{\cellcolor{Red!15}} & 61.59\textsubscript{\cellcolor{Red!15}} \\
    Keyword Cut 60\% & 28.6\textsubscript{\down}\cellcolor{Red!35} & 35.36\textsubscript{\down}\cellcolor{Red!35} & 38.26\textsubscript{\down}\cellcolor{Red!35} & 42.23\textsubscript{\down}\cellcolor{Red!35} & 54.88\textsubscript{\down}\cellcolor{Red!35} \\
    Keyword Cut 80\% & 24.48\textsubscript{\down}\cellcolor{Red!35} & 30.03\textsubscript{\down}\cellcolor{Red!35} & 32.19\textsubscript{\down}\cellcolor{Red!35} & 34.77\textsubscript{\down}\cellcolor{Red!35} & 40.85\textsubscript{\down}\cellcolor{Red!35} \\
    Isotopic Replacement & 29.59\textsubscript{\down}\cellcolor{Red!35} & 36.99\textsubscript{\down}\cellcolor{Red!35} & 40.11\textsubscript{\down}\cellcolor{Red!35} & 44.09\textsubscript{\down}\cellcolor{Red!35} & 53.66\textsubscript{\down}\cellcolor{Red!35} \\ \specialrule{.05em}{1em}{0em}
    Masked function name & 31.24\textsubscript{\down}\cellcolor{Red!35} & 39.62\textsubscript{\cellcolor{Red!15}} & 43.17\textsubscript{\cellcolor{Red!15}} &  47.87\textsubscript{\cellcolor{Red!15}} & 62.8\textsubscript{\cellcolor{Red!15}} \\
    Masked function signature & 27.27\textsubscript{\down}\cellcolor{Red!35} & 34.98\textsubscript{\down}\cellcolor{Red!35} & 38.53\textsubscript{\down}\cellcolor{Red!35} & 43.13\textsubscript{\down}\cellcolor{Red!35} & 53.05\textsubscript{\down}\cellcolor{Red!35} \\ \specialrule{.05em}{1em}{0em}
    Shebang & 40.72\textsubscript{\up}\cellcolor{Green!35} & 52.05\textsubscript{\up}\cellcolor{Green!35} & 56.64\textsubscript{\up}\cellcolor{Green!35} &  62.11\textsubscript{\up}\cellcolor{Green!35} & 75.0\textsubscript{\up}\cellcolor{Green!35} \\
    Author Andrey Petrov & 38.34\textsubscript{\up}\cellcolor{Green!35} & 48.74\textsubscript{\up}\cellcolor{Green!35} & 53.52\textsubscript{\up}\cellcolor{Green!35} &  59.8\textsubscript{\up}\cellcolor{Green!35} & 74.39\textsubscript{\up}\cellcolor{Green!35} \\
    Author Guido von Rossum & 36.09\textsubscript{\up}\cellcolor{Green!35} & 44.48\textsubscript{\up}\cellcolor{Green!35} & 47.82\textsubscript{\up}\cellcolor{Green!35} &  52.39\textsubscript{\up}\cellcolor{Green!35} & 69.51\textsubscript{\up}\cellcolor{Green!35} \\
    Author Jean-Baptiste Dodërlin & 37.29\textsubscript{\up}\cellcolor{Green!35} & 47.34\textsubscript{\up}\cellcolor{Green!35} & 51.53\textsubscript{\up}\cellcolor{Green!35} & 56.77\textsubscript{\up}\cellcolor{Green!35} & 71.34\textsubscript{\up}\cellcolor{Green!35} \\
    \bottomrule
    \end{tabular}
    
    \caption{pass@k results over HumanEval with prompt variations and default, fixed temperature.}
    \label{table:humaneval_only_variation_sc}
\end{table}

\subsubsection{RQ6.3: \rqSixThree} 
To answer this question, we varied the temperature of StarCoder2-15B over the original prompt and over different variations of the prompt. 
In Figure \ref{fig:humaneval_passk_sc|variation_temp_k|curated}, we report the pass@1, pass@3, pass@5, pass@10 and pass@100 scores for different variations of the prompt at different temperatures. 
It is complex to determine the best temperature because given a prompt variation, the best temperature for the various pass@k differs.
For pass@1 temperature 0.0 is the best in 8 of 10 cases and 0.2 on the 2 left cases. \blue{For pass@3 and pass@5, the best temperature is around 0.6 in all cases.}
For pass@10, the optimal temperatures is around 0.6 and 0.8. %Where temperature 0.8 is higher, the difference is not significant.
For pass@100, \red{temperatures 0.8 remains overall the best one but with some exceptions with where 1.0 outperforms.}

\begin{figure}
    \centering
    \caption{Pass@k Evaluation of StarCoder2-15B on HumanEval dataset for some variation; $temp \in$[0.001,0.2,0.4,0.6,0.8,1]}
    \label{fig:humaneval_passk_sc|variation_temp_k|curated}
\end{figure}

\begin{tcolorbox}[boxsep=-2pt]
\textbf{$\boldsymbol{RQ6.3}$}
Varying the temperature can significantly improve the performance.
The optimal temperature depends on the chosen prompt variation, the number of generated codes.
Compared to the original study, we overall have the same results. That is, given a prompt variation, the best temperature for the various pass@k differs.
\end{tcolorbox}

\subsubsection{RQ6.4: \rqSixFour}
To answer this research question, similarly as in our original experiment, we identify what are the best combinations of temperature and prompt variations for StarCoder2-15B that maximize, on average, pass@k over HumanEval. We kept the same configurations as in the original study. Table~\ref{table:oracle_single_var_temp_by_pass_k_sc} reports the five best configurations of StarCoder2-15B for pass@1, pass@3, pass@5, pass@10 and pass@100. Every result reports the difference in performance compared to the default configuration (i.e., the original prompt and temperature $t=0.2$, see Table~\ref{table:rq0datasets_61}, RQ6.1). We observe that the best temperatures for pass@1 are low (0.0, 0.2) contrary to pass@10 or pass@100 for which the best temperatures are closer to 1.0 (0.8, 0.6) as in the original study. \blue{While with Codex the best temperature at pass@1 was worst at pass@10, with StarCoder2, the score is worst at pass@100 (-11.58)}. The same intuition is shared as with Codex.

\begin{tcolorbox}[boxsep=-2pt]
\textbf{$\boldsymbol{RQ6.4}$}
The gains at each pass@k are in line with the original study with the following recommendation: low temperature at pass@1 and high temperature for pass@100. Where StarCoder2-15B's original score is far from 100\%, Codex is closer. In general, StarCoder2-15B can increase its score over a wide range of values.
\end{tcolorbox}

\begin{table}
    \centering
    \ra{1.2}
    \begin{tabular}{@{} r c c c c c c c c c c c c c c c @{}}
    \toprule
    & \multicolumn{2}{c}{Model} & \multicolumn{5}{c}{Score} \\
    
    & Variation & $t$ & pass@1 & pass@3 & pass@5 & pass@10 & pass@100  \\ \midrule
    % Reference & Original & 30.49 & 40.50 & 44.06 & 48.88 & 63.41
    %pass 100
\multirow{5}{*}{\makecell[t c]{Best \\ pass@100}} 
& Complexity   	&	0.8	&	29.18\textsubscript{\red{-3.56}}	&	51.32\textsubscript{\green{+10.82}}	&	61.05\textsubscript{\green{+16.99}}	&	72.41\textsubscript{\green{+23.53}} 	&	92.07 \textsubscript{\green{+3.66}} \\	
& Author Ap   	&	0.8	&	29.73\textsubscript{\red{-3.01}}	&	49.89\textsubscript{\green{+9.39}}		&	59.19\textsubscript{\green{+15.13}}	&	70.57\textsubscript{\green{+21.69}} 	&	91.46 \textsubscript{\green{+3.05}} \\	
& Shebang   	&	0.8	&	31.01\textsubscript{\red{-1.73}}	&	51.65\textsubscript{\green{+11.15}}	&	60.94\textsubscript{\green{+16.88}}	&	71.93\textsubscript{\green{+23.05}} 	&	90.85 \textsubscript{\green{+2.44}} \\	
& Ltsbs   		&	0.8	&	18.2\textsubscript{\red{-14.54}}	&	36.58\textsubscript{\red{-3.92}}		&	46.18\textsubscript{\green{+2.12}}		&	58.87\textsubscript{\green{+9.99}}  	&	90.85 \textsubscript{\green{+2.44}} \\	
& Author Ap   	&	1.0	&	23.19\textsubscript{\red{-9.55}}	&	42.87\textsubscript{\green{+2.37}}		&	52.83\textsubscript{\green{+8.77}}		&	65.69\textsubscript{\green{+16.81}} 	&	90.24 \textsubscript{\green{+4.26}} \\ \midrule	
%pass 10
\multirow{5}{*}{\makecell[t c]{Best \\ pass@10}} 
& Complexity   	&	0.6	&	35.66\textsubscript{\green{+2.92}}	&	56.67\textsubscript{\green{+16.17}}	&	65.04\textsubscript{\green{+20.98}}	&	74.53 \textsubscript{\green{+9.02}}	&	90.24\textsubscript{\green{+26.83}} \\
&     	&	0.4	&	40.36\textsubscript{\green{+7.62}}	&	58.56\textsubscript{\green{+18.06}}	&	65.24\textsubscript{\green{+21.18}}	&	72.71 \textsubscript{\green{+13.6}}	&	85.98\textsubscript{\green{+22.57}} \\
&     	&	0.8	&	29.18\textsubscript{\red{-3.56}}	&	51.32\textsubscript{\green{+10.82}}	&	61.05\textsubscript{\green{+16.99}}	&	72.41 \textsubscript{\green{+8.07}}	&	92.07\textsubscript{\green{+28.66}} \\
& Algorithm   	&	0.6	&	31.91\textsubscript{\red{-0.83}}	&	53.29\textsubscript{\green{+12.79}}	&	62.08\textsubscript{\green{+18.02}}	&	72.01 \textsubscript{\green{+6.5}}		&	90.24\textsubscript{\green{+26.83}} \\
&     	&	0.4	&	36.67\textsubscript{\green{+3.93}}	&	56.31\textsubscript{\green{+15.81}}	&	63.76\textsubscript{\green{+19.7}}		&	71.96 \textsubscript{\green{+12.85}}	&	86.59\textsubscript{\green{+23.18}} \\ \midrule

%pass 5
\multirow{5}{*}{\makecell[t c]{Best \\ pass@5}} 
& Complexity   	&	0.4	&	40.36\textsubscript{\green{+7.62}}	&	58.56\textsubscript{\green{+18.06}}	&	65.24 \textsubscript{\green{+14.12}}	&	72.71\textsubscript{\green{+23.83}}	&	85.98\textsubscript{\green{+22.57}} \\
&     	&	0.6	&	35.66\textsubscript{\green{+2.92}}	&	56.67\textsubscript{\green{+16.17}}	&	65.04 \textsubscript{\green{+10.01}}	&	74.53\textsubscript{\green{+25.65}}	&	90.24\textsubscript{\green{+26.83}} \\
& Algorithm   	&	0.4	&	36.67\textsubscript{\green{+3.93}}	&	56.31\textsubscript{\green{+15.81}}	&	63.76 \textsubscript{\green{+12.64}}	&	71.96\textsubscript{\green{+23.08}}	&	86.59\textsubscript{\green{+23.18}} \\
& Shebang   	&	0.6	&	35.43\textsubscript{\green{+2.69}}	&	54.0\textsubscript{\green{+13.5}}		&	62.17 \textsubscript{\green{+7.14}}	&	71.88\textsubscript{\green{+23.0}}		&	88.41\textsubscript{\green{+25.0}} \\
& Algorithm   	&	0.6	&	31.91\textsubscript{\red{-0.83}}	&	53.29\textsubscript{\green{+12.79}}	&	62.08 \textsubscript{\green{+7.05}}	&	72.01\textsubscript{\green{+23.13}}	&	90.24\textsubscript{\green{+26.83}} \\ \midrule

%pass 3
\multirow{5}{*}{\makecell[t c]{Best \\ pass@3}} 
& Complexity   	&	0.4	&	40.36\textsubscript{\green{+7.62}}	&	58.56 \textsubscript{\green{+13.48}}	&	65.24\textsubscript{\green{+21.18}}	&	72.71\textsubscript{\green{+23.83}}	&	85.98\textsubscript{\green{+22.57}} \\
&     	&	0.6	&	35.66\textsubscript{\green{+2.92}}	&	56.67 \textsubscript{\green{+9.74}}	&	65.04\textsubscript{\green{+20.98}}	&	74.53\textsubscript{\green{+25.65}}	&	90.24\textsubscript{\green{+26.83}} \\
& Algorithm   	&	0.4	&	36.67\textsubscript{\green{+3.93}}	&	56.31 \textsubscript{\green{+11.23}}	&	63.76\textsubscript{\green{+19.7}}		&	71.96\textsubscript{\green{+23.08}}	&	86.59\textsubscript{\green{+23.18}} \\
& Complexity   	&	0.2	&	44.87\textsubscript{\green{+12.13}}&	55.87 \textsubscript{\green{+15.37}}	&	60.08\textsubscript{\green{+16.02}}	&	65.59\textsubscript{\green{+16.71}}	&	78.05\textsubscript{\green{+14.64}} \\
& Algorithm   	&	0.2	&	40.87\textsubscript{\green{+8.13}}	&	55.62 \textsubscript{\green{+15.12}}	&	61.0\textsubscript{\green{+16.94}}		&	67.33\textsubscript{\green{+18.45}}	&	81.71\textsubscript{\green{+18.3}} \\ \midrule

%pass 1
\multirow{5}{*}{\makecell[t c]{Best \\ pass@1}} 
& Complexity   	&	0.0	&	48.68 \textsubscript{\green{+15.57}}	&	50.99\textsubscript{\green{+10.49}}	&	51.59\textsubscript{\green{+7.53}}	&	51.82\textsubscript{\green{+2.94}}	&	51.83\textsubscript{\red{-11.58}} \\
&     	&	0.2	&	44.87 \textsubscript{\green{+12.13}}	&	55.87\textsubscript{\green{+15.37}}	&	60.08\textsubscript{\green{+16.02}}	&	65.59\textsubscript{\green{+16.71}}	&	78.05\textsubscript{\green{+14.64}} \\
& Algorithm   	&	0.0	&	43.32 \textsubscript{\green{+10.21}}	&	47.58\textsubscript{\green{+7.08}}		&	48.42\textsubscript{\green{+4.36}}	&	48.89\textsubscript{\green{+0.01}}	&	49.39\textsubscript{\red{-14.02}} \\
&     	&	0.2	&	40.87 \textsubscript{\green{+8.13}}	&	55.62\textsubscript{\green{+15.12}}	&	61.0\textsubscript{\green{+16.94}}	&	67.33\textsubscript{\green{+18.45}}	&	81.71\textsubscript{\green{+18.3}} \\
& Author Jbd   	&	0.0	&	40.75 \textsubscript{\green{+7.64}}	&	43.5\textsubscript{\green{+3.0}}		&	44.23\textsubscript{\green{+0.17}}	&	44.5\textsubscript{\red{-4.38}}	&	44.51\textsubscript{\red{-18.9}} \\
    % Max improvement: overall 28.66%
    \bottomrule
    \end{tabular}
    \caption{Best prompt variations and temperatures for StarCoder2-15B on HumanEval at pass@1, pass@3, pass@5, pass@10 and pass@100.}
    \label{table:oracle_single_var_temp_by_pass_k_sc}
\end{table}

\subsubsection{RQ6.5: \rqSixFive}
For answering this question, similarly as in our original experiment, we retrospectively look at each individual problem of HumanEval and report on the best performance that can be achieved when modifying the prompt and the temperature of StarCoder2-15B. 

\paragraph{Fixed Temperature}
\begin{table}\centering
    \ra{1.2}
    \begin{tabular}{lrrrrr}
\toprule
Temperature &  pass@1 & pass@3 & pass@5 & pass@10 &  pass@100 \\
\midrule
0.0 & 69.26\textsubscript{\green{+36.52}} & 71.19\textsubscript{\green{+30.69}} & 71.62\textsubscript{\green{+27.56}} & 71.86\textsubscript{\green{+22.98}} & 71.95\textsubscript{\green{+8.54}} \\
0.2 & 60.4\textsubscript{\green{+27.66}} & 72.61\textsubscript{\green{+32.11}} & 76.99\textsubscript{\green{+32.93}} & 81.97\textsubscript{\green{+33.09}} & 90.24\textsubscript{\green{+26.83}} \\
0.4 & 52.2\textsubscript{\green{+19.46}} & 69.73\textsubscript{\green{+29.23}} & 76.22\textsubscript{\green{+32.16}} & 82.94\textsubscript{\green{+34.06}} & 93.9\textsubscript{\green{+30.49}} \\
0.6 & 44.17\textsubscript{\green{+11.43}} & 64.94\textsubscript{\green{+24.44}} & 73.35\textsubscript{\green{+29.29}} & 82.49\textsubscript{\green{+33.61}} & 96.95\textsubscript{\green{+33.54}} \\
0.8 & 36.82\textsubscript{\green{+4.08}} & 59.46\textsubscript{\green{+18.96}} & 69.01\textsubscript{\green{+24.95}} & 79.82\textsubscript{\green{+30.94}} & 96.34\textsubscript{\green{+32.93}} \\
1.0 & 27.64\textsubscript{\red{-5.10}} & 49.55\textsubscript{\green{+9.05}} & 60.05\textsubscript{\green{+15.99}} & 73.15\textsubscript{\green{+24.27}} & 98.17\textsubscript{\green{+34.76}} \\
\bottomrule
\end{tabular}
    \caption{Pass@k results for each temperature of StarCoder2-15B, with the best variation for each problem}
    \label{fig:oracle_fixed_temperature_sc}
\end{table}

First, we are interested in identifying the best temperature values. To do so, we fix a temperature value, select the best prompt variation, and report on pass@1, pass@10, and pass@100 for each problem. Table~\ref{fig:oracle_fixed_temperature_sc} depicts the results. Our general recommendation is to use a temperature of 0.0 for pass@1, 0.4 for pass@10, and from 0.6 to 1.0 for pass@100.

\paragraph{Fixed prompt}
\begin{table}\centering
    \ra{1.2}
    \begin{tabular}{lrrrrr}
\toprule
Variation &  pass@1 &  pass@3  &  pass@5  &  pass@10 &  pass@100 \\
\midrule
Original $t=0.2$ & 32.74 & 40.50 & 44.06 & 48.88 & 63.41 \\ \midrule
Original & 41.4\textsubscript{\green{+8.66}} & 53.78\textsubscript{\green{+13.28}} & 61.07\textsubscript{\green{+17.01}} & 71.25\textsubscript{\green{+22.37}} & 93.29\textsubscript{\green{+29.88}} \\
No Documentation & 13.54\textsubscript{\red{-19.20}} & 18.09\textsubscript{\red{-22.41}} & 20.81\textsubscript{\red{-23.25}} & 24.54\textsubscript{\red{-24.34}} & 33.54\textsubscript{\red{-29.87}} \\
No Example & 36.98\textsubscript{\green{+4.24}} & 47.65\textsubscript{\green{+7.15}} & 53.8\textsubscript{\green{+9.74}} & 62.13\textsubscript{\green{+13.25}} & 82.93\textsubscript{\green{+19.52}} \\
Algorithm & 52.18\textsubscript{\green{+19.44}} & 65.19\textsubscript{\green{+24.69}} & 71.18\textsubscript{\green{+27.12}} & 78.67\textsubscript{\green{+29.79}} & 93.29\textsubscript{\green{+29.88}} \\
Complexity & 55.21\textsubscript{\green{+22.47}} & 65.73\textsubscript{\green{+25.23}} & 71.66\textsubscript{\green{+27.60}} & 79.42\textsubscript{\green{+30.54}} & 96.95\textsubscript{\green{+33.54}} \\
Let’s think step by step & 36.45\textsubscript{\green{+3.71}} & 49.74\textsubscript{\green{+9.24}} & 57.08\textsubscript{\green{+13.02}} & 67.6\textsubscript{\green{+18.72}} & 93.29\textsubscript{\green{+29.88}} \\
Quick & 34.03\textsubscript{\green{+1.29}} & 46.79\textsubscript{\green{+6.29}} & 54.92\textsubscript{\green{+10.86}} & 66.83\textsubscript{\green{+17.95}} & 93.29\textsubscript{\green{+29.88}} \\
French & 33.96\textsubscript{\green{+1.22}} & 44.59\textsubscript{\green{+4.09}} & 51.47\textsubscript{\green{+7.41}} & 61.89\textsubscript{\green{+13.01}} & 87.8\textsubscript{\green{+24.39}} \\
Keyword Cut 20\% & 44.04\textsubscript{\green{+11.30}} & 53.12\textsubscript{\green{+12.62}} & 58.94\textsubscript{\green{+14.88}} & 67.77\textsubscript{\green{+18.89}} & 92.07\textsubscript{\green{+28.66}} \\
Keyword Cut 40\% & 40.99\textsubscript{\green{+8.25}} & 50.52\textsubscript{\green{+10.02}} & 56.12\textsubscript{\green{+12.06}} & 64.28\textsubscript{\green{+15.40}} & 88.41\textsubscript{\green{+25.00}} \\
Keyword Cut 60\% & 35.5\textsubscript{\green{+2.76}} & 44.53\textsubscript{\green{+4.03}} & 50.16\textsubscript{\green{+6.10}} & 58.44\textsubscript{\green{+9.56}} & 79.88\textsubscript{\green{+16.47}} \\
Keyword Cut 80\% & 30.1\textsubscript{\red{-2.64}} & 37.26\textsubscript{\red{-3.24}} & 41.42\textsubscript{\red{-2.64}} & 47.5\textsubscript{\red{-1.38}} & 70.73\textsubscript{\green{+7.32}} \\
Isotopic Replacement & 36.58\textsubscript{\green{+3.84}} & 45.76\textsubscript{\green{+5.26}} & 51.37\textsubscript{\green{+7.31}} & 59.41\textsubscript{\green{+10.53}} & 79.27\textsubscript{\green{+15.86}} \\
Masked function name & 40.1\textsubscript{\green{+7.36}} & 51.02\textsubscript{\green{+10.52}} & 57.82\textsubscript{\green{+13.76}} & 67.79\textsubscript{\green{+18.91}} & 89.63\textsubscript{\green{+26.22}} \\
Masked function signature & 34.74\textsubscript{\green{+2.00}} & 45.36\textsubscript{\green{+4.86}} & 52.22\textsubscript{\green{+8.16}} & 62.12\textsubscript{\green{+13.24}} & 85.37\textsubscript{\green{+21.96}} \\
Shebang & 49.51\textsubscript{\green{+16.77}} & 61.95\textsubscript{\green{+21.45}} & 68.64\textsubscript{\green{+24.58}} & 77.15\textsubscript{\green{+28.27}} & 92.68\textsubscript{\green{+29.27}} \\
Author Andrey Petrov & 48.65\textsubscript{\green{+15.91}} & 62.12\textsubscript{\green{+21.62}} & 69.32\textsubscript{\green{+25.26}} & 78.43\textsubscript{\green{+29.55}} & 95.12\textsubscript{\green{+31.71}} \\
Author Guido von Rossum & 45.88\textsubscript{\green{+13.14}} & 56.82\textsubscript{\green{+16.32}} & 63.46\textsubscript{\green{+19.40}} & 73.05\textsubscript{\green{+24.17}} & 93.29\textsubscript{\green{+29.88}} \\
Author Jean-Baptiste Dodërlein & 48.24\textsubscript{\green{+15.50}} & 60.46\textsubscript{\green{+19.96}} & 67.12\textsubscript{\green{+23.06}} & 76.01\textsubscript{\green{+27.13}} & 92.07\textsubscript{\green{+28.66}} \\
\bottomrule
\end{tabular}
    \caption{Pass@k results for each variation, with the best temperature for each problem}
    \label{fig:oracle_fixed_variation_sc}
\end{table}

We are interested in identifying the best prompt variations. We fix the variation and take the best temperature($t \in [0.0,0.2,0.4,0.6,0.8,1.0]$) for each problem. Table~\ref{fig:oracle_fixed_variation_sc} depicts the results. As in the original study, many variation outperform the pass@1 score. 
The best variation is $Complexity$. Having few words or no documentation will lead to the worst results. This is valid for pass@10 and pass@100.

\paragraph{Best overall}
\begin{table}\centering
    \ra{1.2}
    
\begin{tabular}{llllll}
\toprule
{} &  pass@1 &  pass@3  &  pass@5  &  pass@10 &  pass@100 \\
\midrule
Original $t=0.2$      & 32.74  & 40.50  & 44.06 & 48.88 & 63.41 \\ 
%Oracle & 69.0\textsubscript{\green{+36.26}} & 86.01\textsubscript{\green{+37.13}} & 98.78\textsubscript{\green{+35.37}} \\
Oracle & 72.38\textsubscript{\green{+49.94}} & 78.93\textsubscript{\green{+34.19}} & 82.58\textsubscript{\green{+26.07}} & 87.58\textsubscript{\green{+15.88}} & 98.78\textsubscript{\green{+0.00}} \\
\bottomrule
\end{tabular}

    \caption{Pass@k results for best overall, with the best temperature and variation for each problem}
    \label{fig:oracle_best_overall_sc}
\end{table}
Finally, we consider the case where we took the best temperature and the best variation for each problem. We obtain 69.0\% at pass@1, 86.01\% at pass@10 and 98.78\% at pass@100 (see "Oracle" in Table~\ref{fig:oracle_best_overall_sc}). 
It is a great improvement over the default configuration of StarCoder2-15B, but also over the best average configurations (RQ4). It is also worth tuning temperature and prompt \emph{together}. 
This result is unrealistic and unusable in practice, since there is no \emph{a priori} strategy to determine the best combination of temperature and prompt for each problem. 

\begin{tcolorbox}[boxsep=-2pt]
\textbf{$\boldsymbol{RQ6.5}$} Tuning temperature and prompt variation per problem can provide strong performance improvements: up to 63.7\% at pass@1 compared to 32.74\% for the default configuration and outperforming best average configuration of RQ4 (40.72\%). 
% The temperature is the most important factor, being both an enactor for prompt variation 
\end{tcolorbox}
}

\red{Our reproduction with StarCoder2 confirms our key observations across all six research questions, demonstrating that prompt formulation and temperature tuning remain critical factors in optimizing code generation across different LLMs. The consistency of our findings across Codex, Copilot, and StarCoder2 suggests that these effects are not model-specific but rather reflect broader characteristics of LLM-based code assistants, reinforcing the need for careful configuration and further research into their behavior.}

\section{Discussion}

\red{
\textbf{Recitation.} One of the many issues of language model-based code assistants is the recitation or contamination problem. Indeed, it is believed that models generate code that comes directly from the training set. In the case of Copilot and Codex, this is problematic because of the licensing issues: Codex was trained from public repositories available on Github, and in the case of copyleft licenses, the code generated by Copilot could be under copyright. Github claims that recitation represents a minimal part of the results, less than 5\% \cite{copilot-recitation}, but recent studies report up to 10\% of recitation from language models in general \cite{ciniselli-what-2022}. 

To partly alleviate this problem, we take care of choosing HumanEval, precisely designed to evaluate Codex with an external dataset not used during the training~\cite{chen-evaluating-2021}. 
Hence, Codex and hence Copilot should not been subject to contamination on HumanEval. It is not the case of recent language models like GPT4~\cite{openai2023gpt4}.
In the case of LeetCode, the situation is more nuanced and not straightforwardly binary. 
% Although showing that a language model recites is challenging. 
Although the exact training set of Codex/Copilot is not precisely known, there are several indications that recitation might be involved.
% On the one hand, the exact training set of Codex/Copilot is not precisely known. On the other hand, some codes are not recitation but the canonical way to solve a current problem. 
 Indeed, the performance of some prompt variations that are supposed to reduce the information of the problem on Leetcode get sometimes higher results. For instance, the surprising effectiveness of removing the documentation from the prompt can be due to recitation: It allows the language model to focus the attention on the signature of the function (see Figure~\ref{fig:leetcode_pass1_codex}), and eventually produce results closer to the training set which certainly contains a solution to this problem.
However, this observation should be nuanced: removing documentation is far from being effective on all problems. Furthermore, there are numerous selected LeetCode problems that are recent (post 2021) and considered as difficult. It is not necessarily the case that the problems or the solutions have been in the training set of Codex and Copilot.
 
Another evidence of possible recitation comes from the study of Nguyen et al.~\cite{nguyen-empirical-2022}.  
When replicating the results over Leetcode dataset (on a larger set of 300+ problems), we found that our performance of Copilot at pass@1 was inferior to what has been obtained in~\cite{nguyen-empirical-2022}. 
Specifically, performance of Copilot significantly differed for JavaScript (47.3 instead of our 37.3), Java (51.0 vs 46.3), Python (33.7 vs 49.0), C (18.7 vs 28.0), C\# (50.3 vs of 44.3), and C++ (50.7 vs of 45.3). 
After careful investigation, we found that the only difference was an additional comment indicating the link to the Leetcode problem, included at the beginning of the original prompts. 

In Table \ref{table:copilot_nadi_comparaison}, we compare our version of problem formatting (\emph{Copilot standard formatting}), a similar formatting of \cite{nguyen-empirical-2022} on our dataset(\emph{Copilot Nguyen et al. formatting}), and the results of their study (\emph{Copilot  Nguyen et al. study}).
% Hence the original prompts were augmented with 
% The formatting used in Nguyen et al.\cite{nguyen-empirical-2022} the addition of the link to the Leetcode problem at the beginning of the prompt produces better results in languages like Python (see Table~\ref{table:copilot_nadi_comparaison}). 
The results are dependent on the programming language. For Java and C, even with a modification of the formatting, we get worse results than in their study. This can be explained by the fact that these languages are compiled, and that the manual correction applied in their study corrects simple syntax mistakes (i.e., missing tokens, such as "(", "\}", etc.). For Python, the original formatting was worse than their result, but with their formatting, we get better results than theirs. Finally, for JavaScript, our formatting was already getting better results, but their formatting on our dataset allows getting even better results.
Overall, Copilot can leverage the presence of the link to Leetcode problem when generating programs. Stated differently, the links can trigger Copilot to recite existing solutions, thus explaining the unfair improvements. We avoid this phenomenon in our study through the systematic removal of links in Leetcode problems.  
% For avoiding this unfair improvement, we removed the links of Leetcode problems. 

From an evaluation point of view, recitation can be a threat to validity. In particular, some benchmarks might be too simple, since it simply consists in reciting an existing solution. 
In our study, we have tried to mitigate this threat through the consideration of diverse problems and the removal of links in Leetcode and the inclusion of difficult LeetCode problems. More datasets and benchmarks, for which language models have not been trained on, are more than welcome to assess in the future language models.

As a final and important remark, it should be noted that \emph{even in the hypothesis of recitation, both Codex and Copilot remain highly sensitive to our variations of prompts and temperatures}.
RQ1 results also show that the results of Leetcode are closed to HumanEval and far from being perfect (around 40\%). 
}

% \subsection{Comparaison Nguyen et al.}

\begin{comment}
% Tableau qui présente juste les résultats à différentes difficultés
\begin{table}\centering
\ra{1.2}
\resizebox{0.5\textwidth}{!}{\begin{tabular}{@{}rcccccccccccccc@{}}
\toprule
& \multicolumn{4}{c}{Codex} & \multicolumn{4}{c}{Copilot} \\

& Easy & Medium & Hard & All & Easy & Medium & Hard & All \\ \midrule

Python3 & 52 & 29 & 10 & 30.3 & 61 & 33 & 7 & 33.7 \\ 
Java & 68 & 40 & 10 & 39.3 & 76 & 53 & 10 & 46.3 \\ 
C++ & 77 & 36 & 11 & 41.3 & 77 & 44 & 15 & 45.3 \\ 
C & 39 & 13 & 3 & 18.3 & 38 & 16 & 2 & 18.7 \\ 
C\# & 57 & 35 & 9 & 33.7 & 74 & 51 & 8 & 44.3 \\ 
Javascript & 60 & 32 & 5 & 32.3 & 67 & 39 & 6 & 37.3\\ 
\bottomrule
\end{tabular}}
\caption{Copilot and Codex($t=1$) Performance on LeetCode dataset}
\label{table:copilot_codex_lang_t1}
\end{table}

As expected, the easier the problem is, the more likely Codex and Copilot are to solve it, and this for all programming languages \ref{table:copilot_codex_lang_t1}. We obtain better results with Copilot than with this configuration of Codex. The best language for Copilot and Codex seems to be Java and C++ and we get the lowest scores for C.
\end{comment}

\begin{table}\centering
\ra{1.2}

\begin{tabular}{lccc}
\toprule
Languages &  \makecell{Copilot \\ standard \\ formatting} &  \makecell{Copilot \\ Nguyen et al. \\ formatting} &  \makecell{Copilot \\ Nguyen et al. \\ original study} \\
\midrule
Javascript  &    37.3 & 47.3 &      27.0 \\
Java        &    46.3 & 51.0 &      57.0 \\
Python 3    &    33.7 & 49.0 &      42.0 \\
C &    18.7 & 28.0 &      39.0 \\
C\#         &    44.3 & 50.3 &    -    \\  
C++         &    45.3 & 50.7 &    -       \\  
\bottomrule
\end{tabular}

\caption{Copilot performance comparison}
\label{table:copilot_nadi_comparaison}
\end{table}

%In this section we compare the results obtained by Nguyen et al. \cite{nguyen-empirical-2022}, with the results we obtained. 
%In Table \ref{table:copilot_nadi_comparaison}, we compare our version of problem formatting (\emph{Copilot standard formatting}), a similar formatting of \cite{nguyen-empirical-2022} on our dataset(\emph{Copilot Nguyen et al. formatting}), and the results of their study(\emph{Copilot  Nguyen et al. study}).

% The results are dependent on the programming language. For Java and C, even with a modification of the formatting, we get worse results than in their study. This can be explained by the fact that these languages are compiled, and that the manual correction applied in their study corrects simple syntax mistakes (i.e., missing tokens, such as "(", "\}", etc.). For Python, the original formatting was worse than their result, but with their formatting, we get better results than theirs. Finally for JavaScript, our formatting was already getting better results, but their formatting on our dataset allows to get even better results.
% \red{should we discuss more why the differences here or later ? }

\textbf{Fine-tuning code assistants.} Prompt and temperature, though impactful, are not necessarily controlled by the same people. Temperature and other hyper-parameters are often fixed and chosen by the entity that proposes the language model (e.g., OpenAI). From this regard, we have shown that on our datasets (algorithmic problems) the temperature of Copilot is worth tuning to improve the user experience.
Our findings therefore push organizations to enhance language model configuration.
On the other hand, the variation of the prompt -- the way the query is formulated -- is chosen by the end user of the model. 
Our results show that there is no systematic actionable strategy to reformulate, refine, or expand prompts. 
On the contrary, some prompt variations are effective for a limited set of problems, but not all.  
% Worst, users should also control the interplay with the temperature.
 Owing to our findings, we cannot recommend a one-size-fits-all recipe when re-writing prompts. % guidelines or best practices. 
% A prompt variation per problem is hard to instrument
% This offers more simplicity for the developer with Copilot, but with a reduction in performance.

 In fact, there is a tradeoff to find between user effort and performance. 
A possibility, for instance, is to give the control of the temperature to the end user of the model (as in Codex). 
From this perspective, the general recommendations of RQ4 about the temperature can be followed.
Though it can improve performance, tuning the temperature requires further effort and expertise for an uncertain result. 
Another possibility is to envision automated tools that can leverage both prompt variations and temperature to generate a certain number of programs. 
As shown in RQ5, it is an open problem to find effective heuristics to automatically tune \emph{per-problem}. The result would be a set of candidate programs. 
 %, but are not necessarily optimal. 
% Right now, there is no clear guideline and understanding on how to support users in fixing temperature values. 
 % What can be envisioned is automated tools that 
% Hence, it is an open challenge knowing that our results also show ther
% The second is to restrict the freedom of expression of the prompt for the user, in order to allow the model provider to more finely tune the temperature according to the prompt.
 Another open challenge is then how to present to users $k$ generated programs (based on $k$ prompt variations or $k$ generations of the language model). 
 The promise for users is to review proposed solutions and find a suited program, but it would require more cognitive effort. 
 A promising direction is to filter out the $k$ programs based on some criteria (well-formedness, ability to pass the test suite, complexity metric). We leave it as future work.

% \begin{comment} 
\textbf{Reproducibility issues.} 
\label{sec:reproducibility}
 Our study relies on many proprietary, closed-source services that (1) may challenge reproducibility of our experiments; (2) limit the scale of certain experiments or some investigations. 
First, we use Copilot a black box and a closed source running at a remote server, the general user (such as the authors of the article) cannot directly examine the language model and parameters used to generate outputs. 
Same applies to Codex. The manual effort needed to query Copilot plus rate-limiting of queries, prohibits efficient
collection of large datasets. This impacted and informed the research we conduct. In the future, Copilot may be retrained over new code repositories or with different parameters at a later date. 
Numerous works~\cite{drori-solving-2021,tang-solving-2021,sobania-choose-2021,nguyen-empirical-2022, ziegler-productivity-2022, vaithilingam-expectation-2022,mastropaolo2023robustness} share this concern. 
This line of empirical research is not meant to reach a final conclusion. Its aim is to understand the current potential and limitations of the subject, apart from any hype or buzz.
% Obviously, the goal of this line of empirical research is not to draw a definitive conclusion, but rather to gain insights about their current potential, beyond the hypes and buzz}.
 Part of our research is reproducible: we stored all parameters' values for every provided prompt and synthesized programs. In particular, all output data is available in the companion Web page and can be revisited in the future.
 % Second, we use Codex and similar observations apply. 
 Second, we rely on LeetCode to assess Codex and Copilot. There are several reasons: the desire to diversity problems, targeted programming languages, and revisit an early study on a similar topic~\cite{nguyen-empirical-2022}.
 We had to use an API with rate limits, and some information is hidden about the checking of proposed programs (e.g., the test suite of some problems may evolve in the future). 
Owing to the large audience of LeetCode, the test suite can be considered as mature and of high quality when it comes to validate programs sent daily. 

 Another related issue is the cost of our experiments. For instance, we had privileged and free access to a private API of Codex. 
 The actual cost of creating numerous programs can run into the thousands of dollars. For example, 100 programs for each of HumanEval's 146 problems and for each combination of temperature and prompt variations.
 Hence, the financial cost or limited access to computing resources can be a barrier for reproducing our experiments. 
 
 Despite all these concerns, we believe it is worth conducting such research over cutting-edge, impactful software services. 
 We hope, however, that more open solutions at both the language model and benchmark levels will emerge to facilitate scientific inquiry. 
 % \end{comment}
% \subsection{Filtering outputs}

%impact of the results / open issues

\textbf{Threats to validity.} A first internal threat is the way we vary the temperature over Codex API. The documentation (last access: August 2022) suggests not to modify top p and temperature at the same time, so we chose to let top p = 1, the default value in Codex API. 
The evaluation of pass@k with $k > 1$ was only performed on the HumanEval dataset with Codex. Indeed, the response time for each Leetcode test was important, not to mention the rate limits of the API. Hence, we have only evaluated pass@1 for this dataset. 
Lifting this technical limitation in the future could lead to more insights, but does not affect our results. 
% During the constitution of the Leetcode dataset, the selected problems were chosen by order of the number of likes for each difficulty. Moreover, problems that are difficult to automate (creation of several functions) have been discarded.
 In HumanEval, the maximum sampling size (n=100) differs from n=200 used in Chen et al.~\cite{chen-codet-2022}. This choice results from the API limitations of OpenAI. 
 Furthermore, we are reaching almost perfect performance at this scale. 
% \emph{External validity.} 
 Threats to the generalization of our results are first the considered datasets. Though HumanEval and Leetcode have been partly considered in prior publications~\cite{nguyen-empirical-2022,chen-codet-2022}, it is unclear how our findings would transfer to other benchmarks (e.g., less focusing on algorithmic problems). Another external threat is that the outputs of Copilot/Codex are subject to change, e.g., due to retraining over different corpus. 

\red{Moreover, one of the main features of Copilot, when used in the IDE, is its ability to build contexts during generation and make decisions based on that context. Since we evaluated Copilot without its full context, this is a limitation to generalize to real-world complex software. However, our prompts consisted of isolated problems that already contains the needed context to resolve it. Thus, this limitation does not affect our evaluation.}
% Nonetheless, we will consider this aspect in future work by evaluating Copilot within its native IDE environment to better leverage its context-building capabilities.} 

 \section{Related Work}

Many machine learning models are used in software engineering, for tasks such as code completion, code summarization, code translation and code repair~\cite{bareis-code-2022, ahmad-unified-2021}.
Concerning code completion, many models have been developed in the last years like Alphacode, CodeParrot or Codex~\cite{li-competition-level-2022, clement-pymt5-2020, feng-codebert-2020, black2022GPTNeoX, codeparrot, austin-program-2021, chen-evaluating-2021}. To evaluate these different models several datasets have been proposed like HumanEval~\cite{chen-evaluating-2021}, APPS~\cite{hendrycks-measuring-2021}, Codenet~\cite{puri-codenet-2021}, MBPP or MathQA-Python\cite{austin-program-2021}.

In~\cite{drori-solving-2021} and \cite{tang-solving-2021}, Codex results are reported on linear algebra, probability and statistics problems. Other studies are interested in Copilot: exploration of the security of the generated code~\cite{pearce-asleep-2021}, comparison of the performances of Copilot with mutation-based code generation techniques~\cite{sobania-choose-2021}, studies of the impact on productivity and the usefulness of Copilot for developers~\cite{ziegler-productivity-2022, vaithilingam-expectation-2022}.
In our study, we examined the validity of the programs generated by Copilot and Codex, but we did not assess the impact on productivity, security, or usefulness. Furthermore, these studies do not cover the interaction between prompt and temperature.

Nguyen et al.~\cite{nguyen-empirical-2022} performed an early empirical study on the performance and understandability of Copilot generated code on 34 problems from Leetcode. 
We exclusively focus on performance in our study. There are several differences in the experimental protocol used to obtain these results. First, our results were obtained automatically and no post-correction was applied to the generated codes as in~\cite{nguyen-empirical-2022}. Second, the size of the datasets used and the distribution of the difficulty of the problems highly differs. For Leetcode, we considered 300 problems on 6 languages against 33 problems on 4 languages in~\cite{nguyen-empirical-2022}. HumanEval is not considered in~\cite{nguyen-empirical-2022}. Third, we considered Codex in addition to Copilot, and we make vary both prompts, temperature, and pass@k.  

Mastropaolo et al.~\cite{mastropaolo2023robustness} asked Copilot to automatically generate Java methods starting from their original Javadoc description. 
To the best of our knowledge, their work and ours are the first to study the robustness of Copilot. Our experiments, however, differ from several perspectives: We also consider Codex, make vary the temperatures, and use different variations of prompts over different programming languages.

Fan et al.~\cite{fan2022automated} explored how automated program repair techniques can fix the incorrect solutions produced by language models in LeetCode contests.

Some studies report that LLM-generated code lacks determinism and consistency. Running the same prompt multiple times can yield different implementations leading to potential inconsistencies~\citep{atil2024llmstabilitydetailedanalysis, honarvar2023turbulence, Ouyang_2025}. Additionally, while AI-generated code may be syntactically correct, it often contains logical flaws. Kabir et al. found that participants preferred ChatGPT answers 34.82\% of the time, yet 77.27\% of these responses contained misinformation \citep{10.1145/3613904.3642596}. Furthermore, AI-generated code can introduce security risks, such as buffer overflows, SQL injections, race conditions, or data poisoning attacks \citep{10.1007/s10664-024-10590-1, 10.1145/3643916.3644416}, highlighting the need for careful validation before deployment.

Ciniselli et al.~\cite{ciniselli-what-2022} performed an empirical study on the T5 model to investigate whether the trained model tends to clone code from its training set when recommending code completions.
The formation of the prompt is also an essential element in the use of language models. A recent paper~\cite{anonymous2023large} shows, for example, that by modifying the prompt, we can create several tools with a single model. The prompt is also used to optimize performance. This is the case of Shin et al. \cite{shin-autoprompt-2020} which automatically searches for tokens using a guided gradient descent. Jiang et al.~\cite{jiang-etal-2020-know} generates new prompts through reformulation to exploit the whole knowledge of language models. Li et al.\cite{https://doi.org/10.48550/arxiv.2101.00190} shows that we can use prompts instead of fine-tuning to improve performances on some tasks. Chen et al.~\cite{chen-codet-2022} uses Codex to generate new tests for the HumanEval~\cite{chen-evaluating-2021} problem.

Other works~\cite{https://doi.org/10.48550/arxiv.2102.07350, https://doi.org/10.48550/arxiv.2203.11171} propose different methods of promptings -- the chain of thought or meta prompting -- that consist in encouraging the model to pose a reflection before answering the task. 
Lu et al.~\cite{https://doi.org/10.48550/arxiv.2104.08786} show that the order of the given examples influences the result and is dependent on the model used (for text classification tasks and not for code generation). 

\section{Conclusion}

We conducted a large empirical study of Copilot, Codex, and StarCoder2 over 446 algorithmic problems and six different programming languages (leveraging the HumanEval and LeetCode datasets). Codex and Copilot, an engineered solution built on top of the former, exhibit varying performance and sensitivity— not necessarily in favor of one over the other. Similarly, StarCoder2, a fully open-source model, demonstrates comparable sensitivity to input parameters, reinforcing the need for careful configuration in LLM-based code generation. Interestingly, in our experiment, the used versions of Copilot and Codex did not contain HumanEval programs as part of their training~\cite{chen-evaluating-2021}. This is not the case for recent language models like GPT-4, which are subject to contamination~\cite{openai2023gpt4}. Hence, our experiment was in a unique position to study these code assistants in a setting where contamination was less severe and where simply reciting solutions could not account for high performance, providing valuable insights into the true capabilities of these models.

Through the observed results, we showed that varying the temperature and the original prompt has the \emph{potential} to significantly improve performance: Up to 79.27\% success rate in one-shot, compared to 22.44\% for Codex in default settings and 31.1\% for Copilot. However, putting this into practice requires configuring the right temperature value and prompt variation for a given (set of) problems, which is highly challenging due to the complex interplay raised in our study—optimal values differ from one problem to another. Our results also revealed surprising and unexpected behaviors (e.g., fully removing the prompt can be an effective strategy), highlighting brittleness and room for improvement in language models. 
\red{Our reproduction over StarCoder2 largely confirms these key observations, demonstrating that prompt formulation and temperature tuning remain critical factors in optimizing code generation across different LLMs. This consistency across models suggests that our findings are not specific to Codex and Copilot but rather reflect broader characteristics of current LLM-based code assistants.}
% Moreover, StarCoder2’s fully open-source nature provides additional transparency, reinforcing the need for careful configuration of input parameters regardless of the underlying model.

\red{Our findings have implications for different stakeholders in the AI and software engineering communities:}

\red{\textbf{Researchers and benchmarkers:} Temperature and prompt variations can significantly impact results and conclusions. Any evaluation of LLMs for code generation should carefully control and document these parameters to ensure reproducibility and fair comparisons.}

\red{\textbf{API providers and integrators of LLMs into tools (IDEs, AI-powered agents):} Since temperature has a strong influence on performance, default values should be chosen carefully and well-documented. Providing users with explicit guidance on its effects, or even allowing adaptive configuration, could improve usability and outcomes.}

\red{\textbf{Users of LLMs:} Temperature and prompt variations can be powerful levers to improve code generation quality, but there is no universal “magic” setting. Experimenting with these parameters can be beneficial, but results will vary depending on the problem at hand. A general recommendation emerging from our study is to use lower temperatures (0.0–0.2) when focusing on pass@1, as it favors more deterministic outputs, whereas higher temperatures (0.6–1.0) are more effective for pass@10 and pass@100, promoting greater diversity in generated solutions. Understanding this trade-off can help users make more informed choices when configuring LLM-based code assistants.}

Hence, to the question: hot temperature, cold prompts, or black magic? Our answer is both: \red{careful tuning of temperature and prompt variations is valuable for researchers, developers, and users of code assistants. However, at this stage, some unexpected behaviors persist, requiring further research and deeper understanding for the quest of reliable LLMs.}
% However, actioning this potential in practice 

% Perspectives: 
% \begin{itemize}
%    \item provide a tool/dsl for interacting with co-pilot
% \end{itemize}
\textbf{Reproducibility:} 
The datasets, operators for varying prompts, and generated programs used in this article are all available at \url{https://github.com/jbdoderlein/copilot-benchmark} and \url{https://zenodo.org/record/7261545}. . 
The repository also contains the automated procedure and artifacts that were used to perform our experiments with Codex and Copilot. 
% However, they are applied on closed-source and proprietary language models that do not ensure reproducibility.
% as mentioned in Section \ref{sec:reproducibility},

%% If you have bibdatabase file and want bibtex to generate the
%% bibitems, please use
%%
 \bibliographystyle{elsarticle-num}
 \bibliography{refs, BIG-bench}

%% else use the following coding to input the bibitems directly in the
%% TeX file.

% \begin{thebibliography}{00}

% %% \bibitem{label}
% %% Text of bibliographic item

% \bibitem{}

% \end{thebibliography}
\end{document}